\documentclass[a4paper,11pt]{article}
\pdfoutput=1 

\usepackage{jcappub} 

\usepackage[T1]{fontenc} 

\usepackage{amsmath,amssymb,epsfig}
\usepackage{wrapfig}
\usepackage{gensymb}
\usepackage[english]{babel}
\usepackage{graphics}
\usepackage{subcaption}
\usepackage{hyperref}
\usepackage[utf8]{inputenc}

\usepackage{booktabs}

\newcommand{\be}{\begin{equation}}
\newcommand{\ee}{\end{equation}}

\newcommand{\diff}{\text{d}}

\newcommand{\mc}{\multicolumn}

\title{High Energy Cosmic Rays from Fanaroff-Riley Radio Galaxies}

\author[a, 1]{B. Eichmann,\note{Corresponding author.}}

\affiliation[a]{Ruhr Astroparticle and Plasma Physics Center (RAPP Center), Ruhr-Universit\"at Bochum, Institut f\"ur Theoretische Physik IV/ Plasma-Astroteilchenphysik, 44780 Bochum, Germany}

\emailAdd{eiche@tp4.rub.de}

\abstract{
The extended jet structures of radio galaxies (RGs) represent an ideal acceleration site for High Energy Cosmic Rays 
(HECRs) and a recent model showed that the HECR data can be explained by these sources, if the arrival 
directions of HECRs at energies $\lesssim 8\,\text{EeV}$ from a certain RG, Cygnus A, are isotropized.

First, this work introduces the inverted simulation setup in order to probe the isotropy assumption. 
Here, different extragalactic magnetic field models are compared showing that either a magnetic 
field of primordial origin that yields a high field strength in the large scale structures of the Universe is 
needed, or a significant contribution by a multitude of isotropically distributed sources. 

Secondly, the HECRs contribution by the bulk of RGs of different Fanaroff-Riley (FR) type is determined. 
Here, the most recent FR-type dependent radio-to-CR correlations $Q_{\rm cr}\propto L_{\rm 
radio}^{\beta_L}$ are used, and the impact of the slope $\beta_L$ on the HECRs is analyzed in detail. 
Finally, it is carved out that FR-II RGs provide a promising spectral behavior at the hardening part of the CR flux, 
between about $3\,\text{EeV}$ and $30\,\text{EeV}$, but most likely not enough CR power. 
At these energies, FR-I RGs can only provide an appropriate flux in the case of a high acceleration 
efficiency and $\beta_L\gtrsim 0.9$, otherwise these sources rather contribute below $3\,\text{EeV}$. Further, the 
required acceleration efficiency for a significant HECR contribution is exposed dependent on $\beta_L$ and the CR 
spectrum at the acceleration site.
}


\begin{document}
\maketitle
\flushbottom

\section{Introduction}
The origin of the High Energy Cosmic Rays (HECRs) is still one of the great enigmas of modern astrophysics. 
HECRs are defined as fully ionized nuclei that penetrate Earth's atmosphere with an energy above a 
few PeV --- the so-called knee in the energy spectrum --- including the most energetic ones, the so-called Ultra High 
Energy Cosmic Rays (UHECRs) that provide energies above a few EeV --- the so-called ankle in the energy spectrum.
From observatories like the Pierre Auger Observatory (Auger) and the Telescope Array (TA) experiment at the 
highest energies as well as KASCADE, KASCADE-Grande and a few other detectors at lower energies, there are basically 
three main observational characteristics, that describe our current knowledge of the HECRs:
\begin{enumerate}
 \item[(1.)] The energy spectrum, which changes at about $0.4\,\text{EeV}$ --- the so-called second knee --- to a 
stepper power-law distribution with a spectral index of about 3.3 and flattens above the ankle at about $3\,\text{EeV}$ 
to a spectral index of 2.6 and a sharp flux suppression above about $30\,\text{EeV}$ \cite{0954-3899-34-10-R01, 
Abbasi:2007sv, 2010PhLB..685..239A}. 
 \item[(2.)] The chemical composition, that shows a decrease of the fraction of heavier elements between about
$0.1\,\text{EeV}$ and $2\,\text{EeV}$, changing to an increase at energies $>2\,\text{EeV}$ 
\cite{Abraham:2010yv, KAMPERT2012660, Aab:2014aea, ObservatoryMichaelUngerforthePierreAuger:2017fhr}. 
 \item[(3.)] The arrival directions, that are usually expressed in terms of the multipoles of their spherical 
harmonics. Between about $0.01\,\text{EeV}$ and $8\,\text{EeV}$ there are no significant hints of anisotropy 
\cite{0004-637X-870-2-91}. However, at higher energies Auger recently reported a $5\sigma$ detection of a dipole with 
an amplitude of $\approx 6.5\%$, while higher-order multipoles are still consistent with isotropy \cite{Aab:2017tyv}. 
Further analysis showed a significant modulation of the first harmonic in right ascension above 
$8\,\text{EeV}$, as well as an increase of its amplitude above $4\,\text{EeV}$ \cite{Aab_2018}.
\end{enumerate}
A likely source candidate of those extremely energetic particles are radio galaxies (RGs) due to their powerful 
acceleration sites within the jets, as already noted by Hillas in 1984 \cite{Hillas:1985is}. In particular the shocks 
caused by the backflowing material in the lobes of RGs represent an ideal acceleration site for HECRs 
\cite{doi:10.1093_mnras_sty2936}. Fanaroff and Riley classified two major types of radio galaxies 
\cite{1974MNRAS.167P..31F}: FR-I RGs, in which the jets are terminating within the galactic environment on scales of a 
few kiloparsec, so that the brightness decreases with increasing distance from the central object; and FR-II RGs, 
where the jets extend on scales of $\gtrsim 100\,$kpc deep into extragalactic space causing an increased brightness 
with distance.  
This morphology distinction obviously correlates with radio power, so that sources with $L_{178} \lesssim 
2\,{\times}\,10^{25}\,\text{W}\,\text{Hz}^{-1}\,\text{sr}^{-1}$ tend to be FR-I galaxies, while sources with $L_{178} 
\gtrsim 2\,{\times}\,10^{25}\,\text{W}\,\text{Hz}^{-1}\,\text{sr}^{-1}$ usually have a FR-II morphology.\footnote{There 
are notable exceptions to this radio power distinction, like the very powerful FR-I galaxy Hydra A.}

In a recent study \cite{1475-7516-2018-02-036} --- hereafter referred to as E+18 --- it is shown that all of the 
observational characteristics of UHECRs can be explained by Centaurus A, a FR-I RG, and Cygnus A, a FR-II RG, if the 
arrival directions of the light CRs from Cygnus A get isotropized due to significant deflections by the 
extragalactic magnetic field (EGMF), providing a rms deflection of $\theta_{\text{rms}}\gtrsim 25\degree 
(\bar{E}/100\,\text{EeV})^{-1}$ for a mean energy $\bar{E}$ of the propagating CR. 
Note that a single source cannot provide an isotropic distribution on a finite observer sphere, as the 
corresponding phase space distribution cannot be homogeneous. But if the CRs from Cygnus A lack their initial 
directional information, i.e.\ $\theta_{\text{rms}}\sim 360\degree$, it is expected according to E+18 that the total CR 
contribution by Cygnus A and Centaurus A, which are located at almost opposite directions, 
provides a good agreement with the dipole strength at $E\sim 8\,\text{EeV}$.
Thus, the dipole is predominantly caused by Centaurus A, which is accidentally located at about the same direction 
as the dipole. However, it needs a heavy ejecta in order to provide enough deflections by 
the Galactic magnetic field\footnote{Here, the GMF model of Jansson \& Farrar \cite{2012ApJ...757...14J} is adopted.} 
to obtain an accurate fit to the observed dipole amplitude. Such an ejecta is motivated by interactions within 
the galactic environment, e.g.\ gaseous shells \cite{2010ApJ...720L.155G}, where the jet typically dissipates in the 
case of FR-I RGs. In order to avoid a heavy jet scenario another significant contribution above $8\,\text{EeV}$ is 
needed, that only M87 and Fornax A can provide according to the E+18 model. Due to the lack of reliable arrival 
directions of events from Cygnus A, the authors do not investigate any additional characteristics of the dipole or 
constrain the contributions by M87 and Fornax A. However, these individual RGs can only explain the UHECR data, if 
about $37\%$ of the CRs from Cygnus A around $8\,\text{EeV}$ are deflected by more than $\sim 300\degree$, hence, the 
initial CR momentum needs to be isotropized.

Unfortunately, the magnetic fields in extragalactic space and our Galaxy are poorly known, and one of the most 
sophisticated descriptions of the EGMF, given by Dolag et al.~\cite{2005JCAP...01..009D} --- hereafter referred as 
D+05, is constrained to a maximal distance of about 120 Mpc. Therefore, the E+18 model does not included 
the impact of deflections on the CRs from Cygnus A at a distance of about $255\,\text{Mpc}$ \cite{2012A&A...544A..18V} 
and a reliable test of the isotropy assumption is missing so far. Further, it needs to be taken into 
account, that Hackstein et al.~\cite{10.1093/mnras/stx3354} --- hereafter referred as H+18 --- have expanded the D+05 
model by introducing the most recent initial conditions of Sorce et al.\ \cite{10.1093/mnras/stv2407}, and provide a 
set of different EGMF models within a volume of $(500\,\text{Mpc})^3$. There are two basic types of EGMF scenarios: The 
primordial models with three different seed field assumptions; and the astrophysical models with three different energy 
budget assumptions, where the impulsive thermal and magnetic feedback in haloes generates the EGMF. Note, that all of 
these different models are constrained by the local observational data. Nevertheless, the cumulative filling 
factors\footnote{The filling factor indicates the fraction of the total volume filled with magnetic fields higher than 
a certain reference value.} of the different EGMF models differ significantly: In principle, the primordial H+18 
models show the highest cumulative filling factors, followed by the astrophysical H+18 models and the D+05 model. Only 
at field strengths $\lesssim 10^{-12}\,\text{G}$ the D+05 model provides a higher filling factor than the 
astrophysical H+18 models.

Another important outcome from E+18 has been the subdominance of the UHECR flux by the non-local RG population, i.e.\ 
the mean contribution from RGs beyond 120 Mpc, above the ankle. But, its spectral behavior has indicated that a 
significant contribution below the ankle is still possible. In addition, the description of the average non-local RG 
population has not differentiated between FR-I and FR-II types, however, FR type dependent radio luminosity functions 
\cite{doi:10.1046_j.1365-8711.2001.04101.x} and radio luminosity to jet power correlations \cite{0004-637X-767-1-12, 
doi:10.1093_mnras_stv2712} indicate the need for a more detailed investigation of the average HECR contribution from 
RGs.

The paper is organized as follows: In Sect.~2 the simulation setup is introduced that provides an estimate of the mean 
deflections of HECRs from Cygnus A in the EGMF model of D+05 as well as the models of H+18, and 
subsequently the ``isotropy assumption'' is probed. In Sect.~3 the continuous source function of HECRs is reinvestigated 
and the average contributions by the bulk of FR-I and FR-II RGs to the observed HECR flux is constrained. All 
simulations are carried out with the publicly available code CRPropa3 \cite{1475-7516-2016-05-038}.
\section{UHECRs from Cygnus A}
\label{Sec:CygA}
To estimate the EGMF effect on HECRs from Cygnus A, a magnetic field structure up to at least $255\,\text{Mpc}$ is 
needed as well as an efficient propagation algorithm to obtain sufficient statistics. 

\subsection{Inverted simulation setup}
\label{Sec:invertedSetup}
Due to the lack of reliable large-scale EGMF structures, the inner cube of the D+05 field, with an edge length 
$l_D\simeq 170 \,\text{Mpc}$, is used and continued reflectively at its boundaries.
Hence, also the extended EGMF stays divergency free. 
In a similar manner also the H+18 models are continued up to necessary extension, as the available 
models\footnote{\url{https://crpropa.desy.de/}} are limited to a volume of $(250\,\text{Mpc})^3$.

\begin{figure}[tbh]
  \centering
    \includegraphics[width=0.6\textwidth]{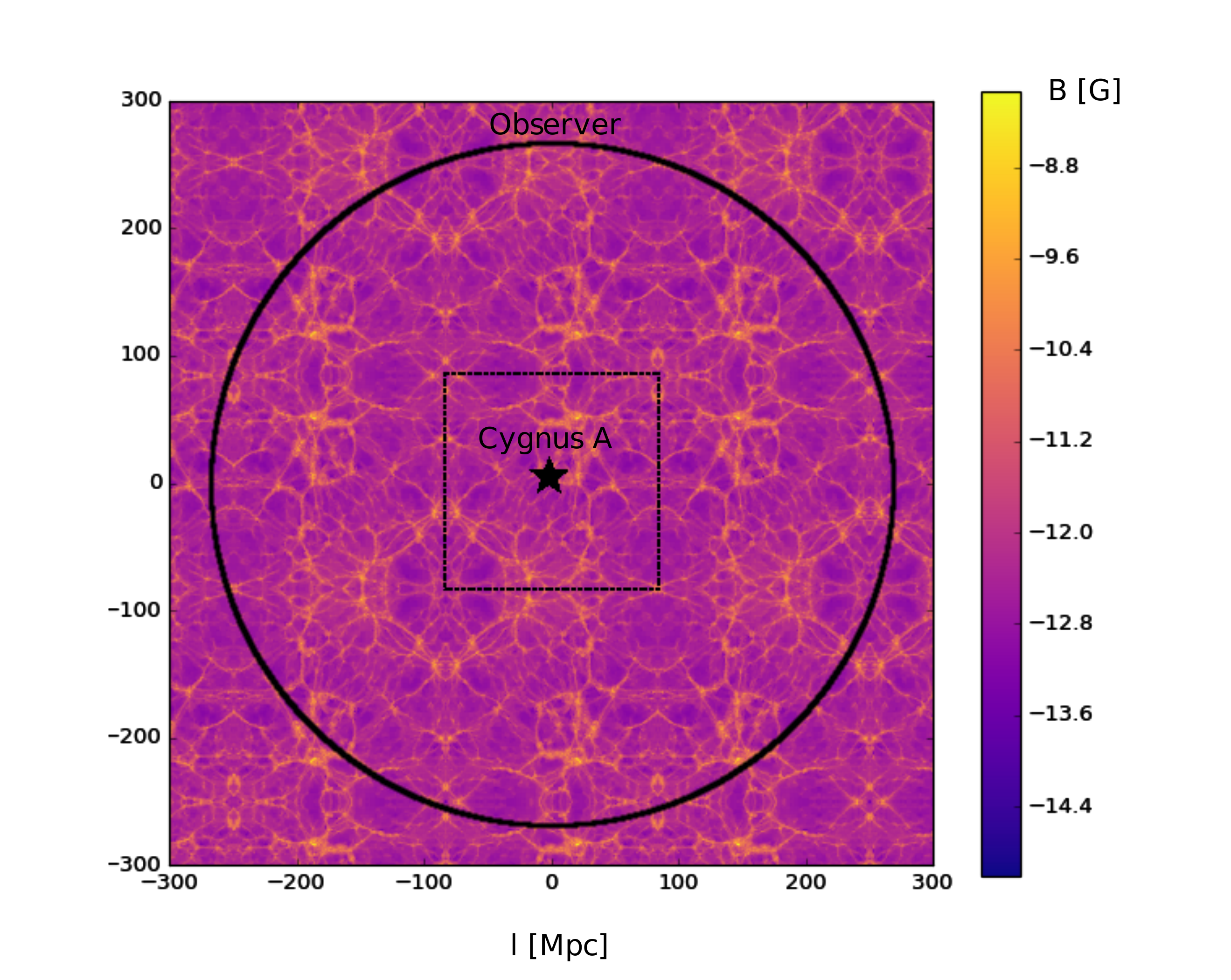}
\caption{Sketch of the inverted simulation setup with the extended EGMF. The thin dashed line marks 
the inner cube of the original D+05 field.}
\label{BfieldSketch}
\end{figure}

Subsequently, an inverted simulation setup is used, where the source is placed at the center of an observer sphere, 
whose radius is determined by the distance of Cygnus A as sketched in the left Fig.\ \ref{BfieldSketch}. 
All CR candidates that cross the spherical surface are collected, but kept in the simulation. So, even the 
proper arrival directions can be estimated by using the zenith angle, as well as the proper spatial positions of the 
Earth and the source. For more details on the reconstruction of the proper arrival directions as well as a discussion 
on the corresponding uncertainties the reader is referred to Appendix \ref{invertedDetails}.  

In principle, the huge benefit of the inverted simulation setup is the 
significant gain of statistics --- with respect to the regular simulation setup used in E+18, since all ejected 
particles will reach the observer, if no additional constrains are used that reject particles from the 
simulation. In the following, the impact of energy losses is not taken into account and a maximal trajectory length of 
$5000\,\text{Mpc}$ is used\footnote{The Hubble time constraints the maximal trajectory length to about 
$4423\,\text{Mpc}$.}.
However, this simulation method is obviously at the expanse of an EGMF structure that is able to represent the proper 
spatial distribution in the local Universe. 
But, in the case of large scale propagations as well as the absence of extragalactic lenses close to the 
Earth, it is expected that the impact of the EGMF is determined by its large scale properties. Hence, the deflection 
rather depends on the distance to the source than on its certain spatial position. 
In the following, 30 arbitrary source positions within the EGMF structure are elaborated in order to avoid 
the impact of the chosen spatial setting. For each setting, $10^5$ individual CR candidates with a fixed rigidity $R\in 
[50\,\text{PV},\,\,1000\,\text{EV}]$ are simulated providing a mean deflection $\bar{\theta}$ and a mean trajectory 
length $\bar{l}_{\rm traj}$. Note, that the deflection angle $\theta_i$ of individual candidates are evaluated using 
the angle between the detected momentum of the CR candidate and the source-to-point-of-detection vector, so that in the 
case of large deflections, that cause an isotropization of the final momenta, $\bar{\theta}$ converges towards 
$90\degree$. 
\subsection{Mean HECR deflection}
\label{Sec:arrDir}
If Cygnus A and Centaurus A are the dominant UHECR sources, as suggested by the E+18 model, the arrival directions 
of CRs from Cygnus A need to be almost isotropically distributed at $E\sim 8\,\text{EeV}$. Thus, the mean deflection needs to be converged towards 
$\bar{\theta}_{\text{iso}}\sim 90\degree$. 
\begin{figure}[tbh]
  \centering
    \includegraphics[width=0.49\textwidth]{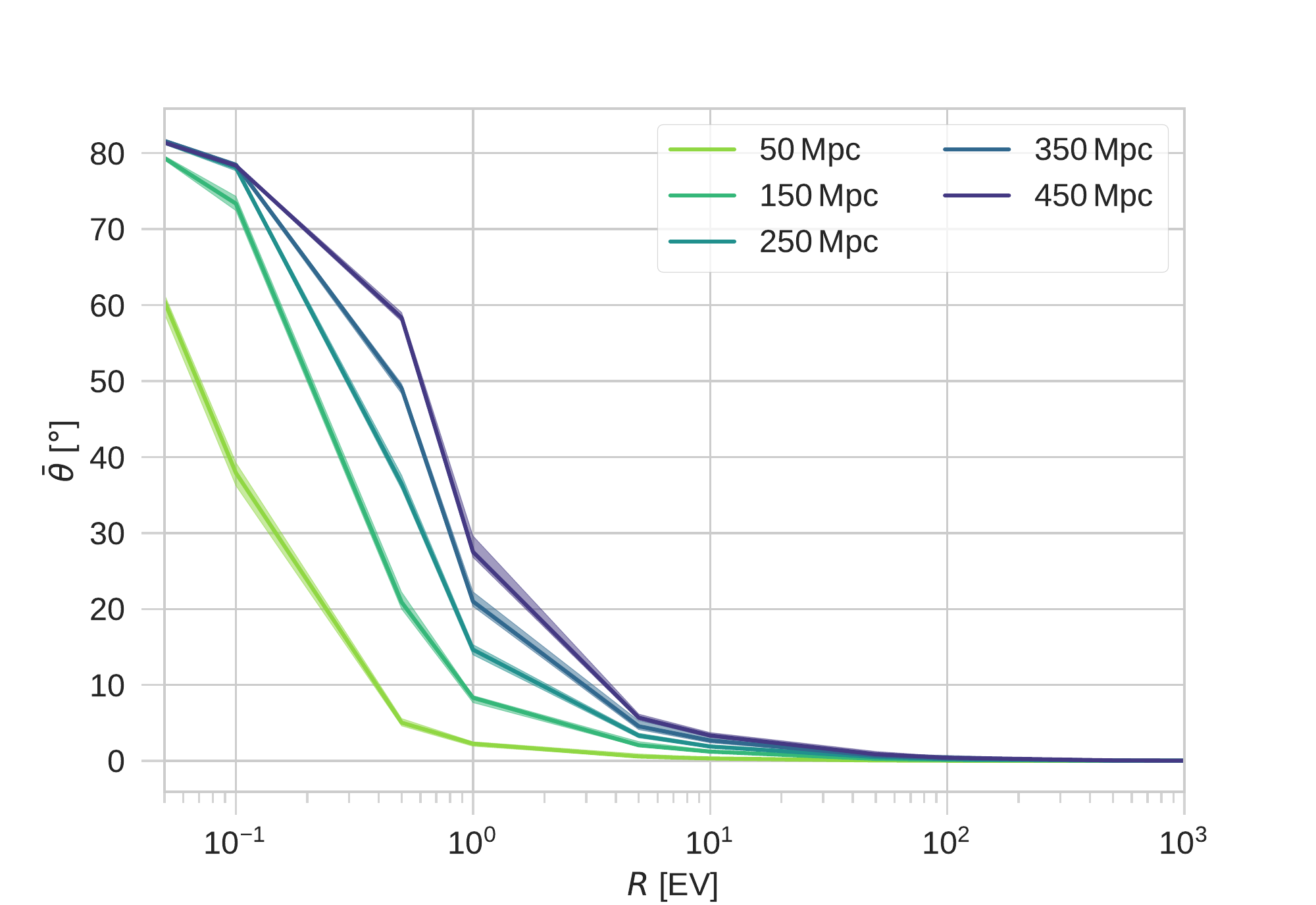}
    \includegraphics[width=0.49\textwidth]{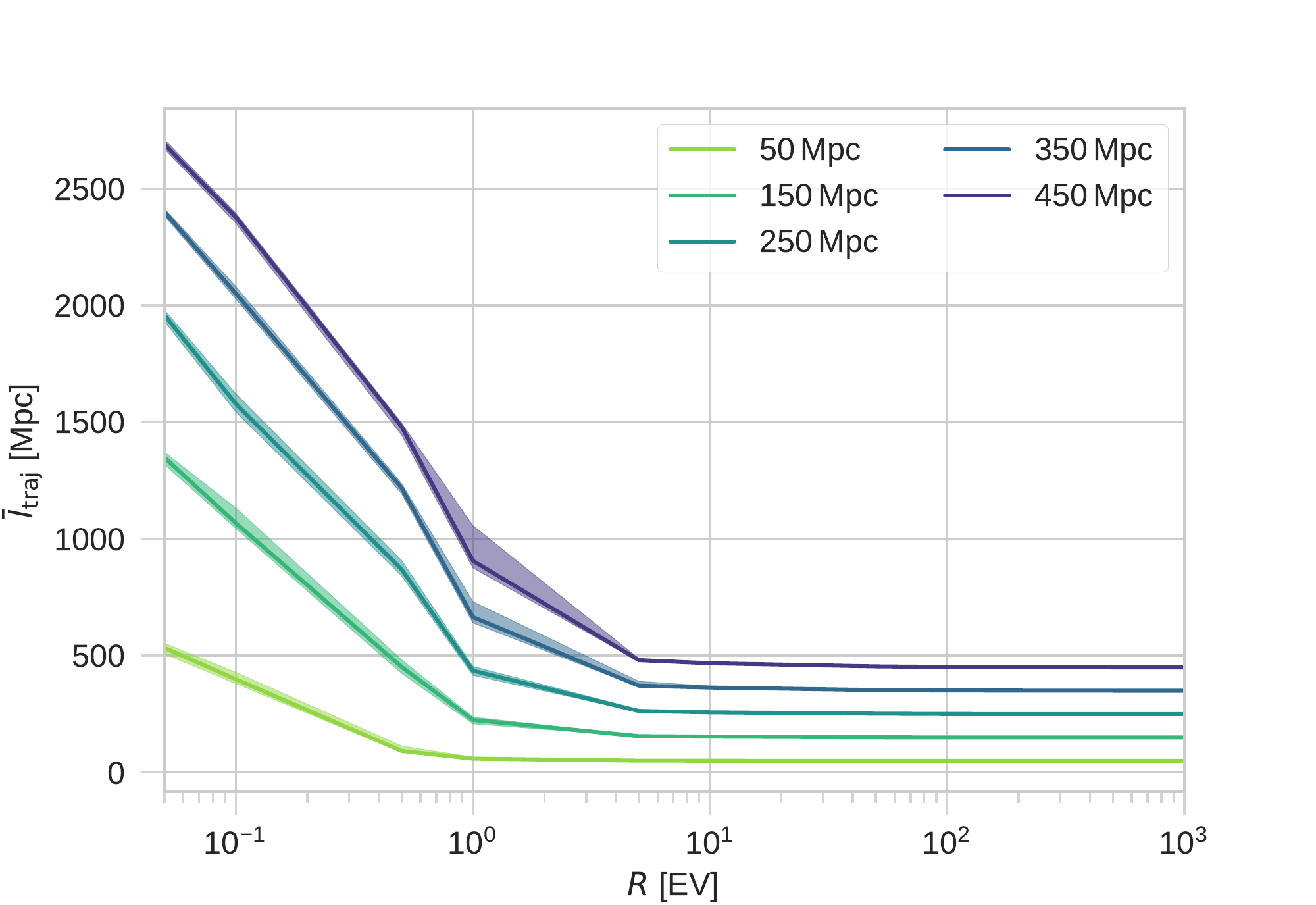}
\caption{Mean deflections (\emph{left}) and mean trajectory lengths (\emph{right}) of CRs dependent on rigidity in the 
extended D+05 field for different source distances. The shaded band indicates the uncertainty due to the arbitrary 
spatial position of the sources.}
\label{rmsDefl_dolag}
\end{figure} 

The left Fig.~\ref{rmsDefl_dolag} displays that even at $R\sim 0.1\,\text{EV}$, i.e.\ a 
high charge numbers like $Z \sim 26$ for energies above the ankle, the ejected CRs by Cygnus A are not completely 
isotropized by the EGMF of D+05. 
In addition, such a heavy CR contribution by Cygnus A can clearly be ruled out, as heavy nuclei suffer from 
photo-disintegration, so that 
the CRs can hardly keep such a high charge number while propagating to Earth. Further, an iron dominated ejecta cannot 
be motivated physically.
In the case of a light CR ejecta, i.e. solar like abundances, even source distances of several hundreds of Mpc yield 
not enough UHECR deflections by the extended D+05 magnetic field to obtain an agreement with the observed 
dipole amplitudes. Further, the resulting mean trajectory lengths $\bar{l}_{\rm traj}$ almost equals the source distance at rigidities
$R\gtrsim 5\,\text{EV}$ as shown by the right Fig.~\ref{rmsDefl_dolag}.

\begin{figure}[tbh]
  \centering
    \includegraphics[width=0.49\textwidth]{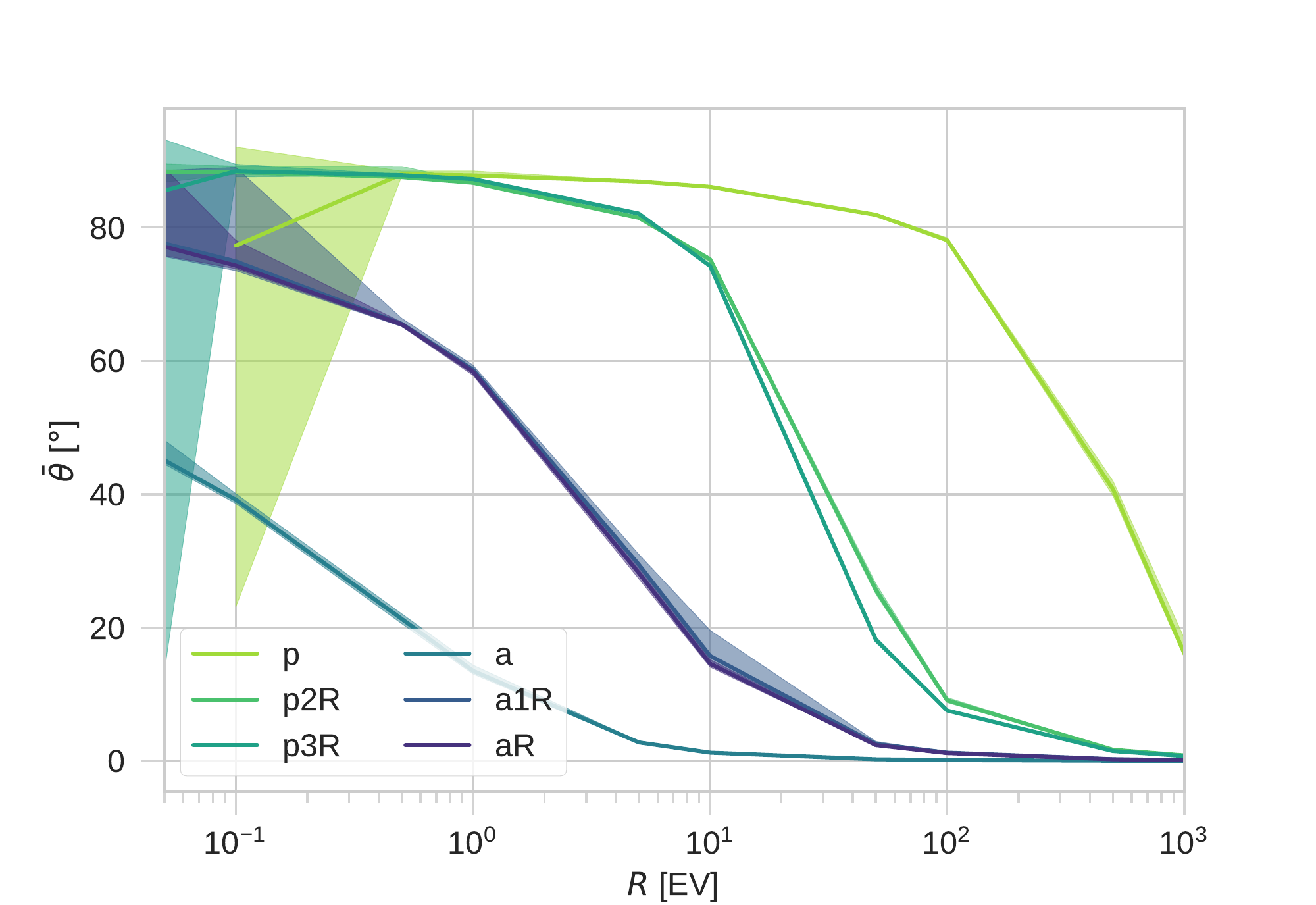}
    \includegraphics[width=0.49\textwidth]{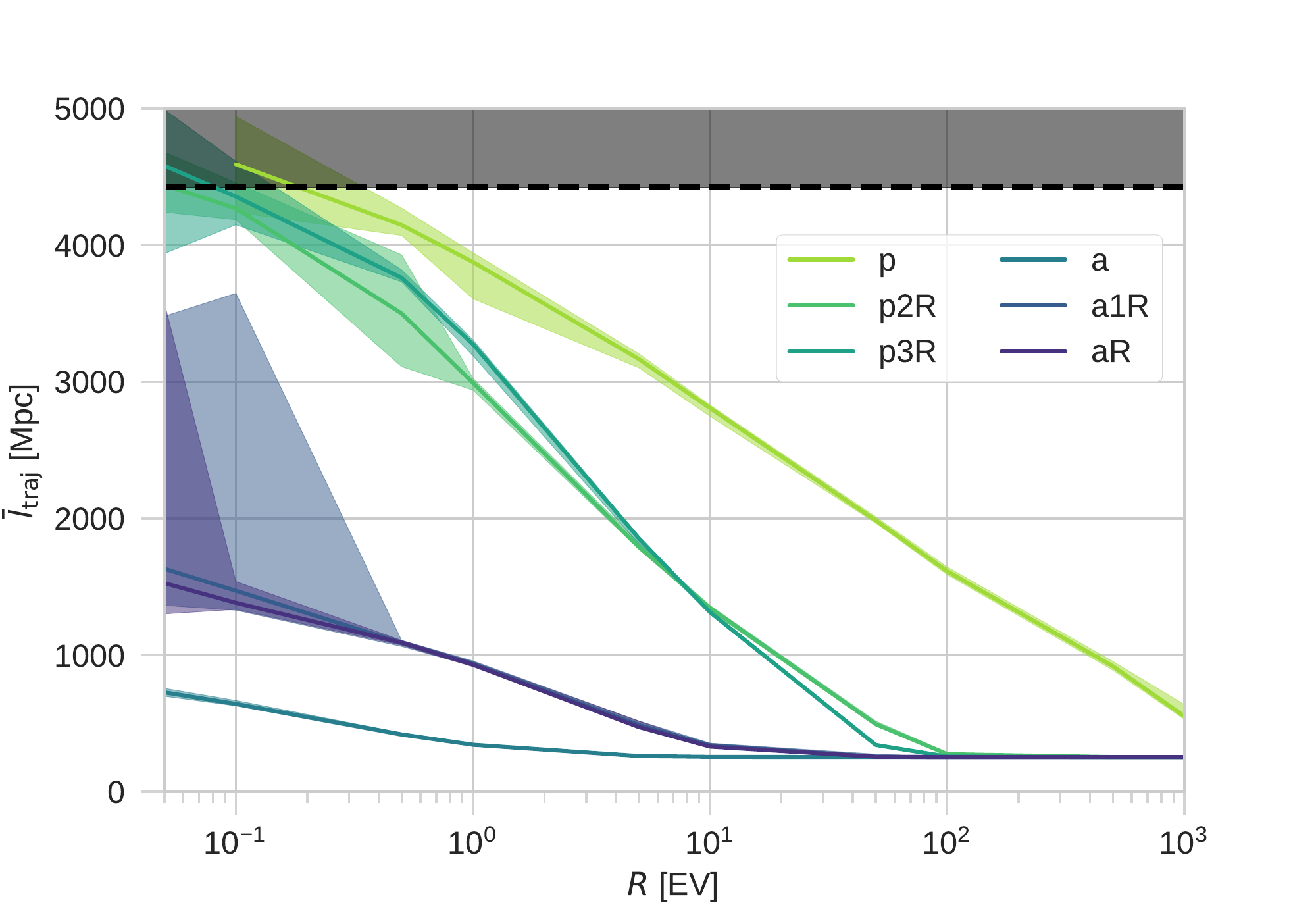}
\caption{Mean deflections (\emph{left}) and mean trajectory lengths (\emph{right}) of CRs dependent on rigidity in the 
different H+18 EGMF models --- 'p' denotes primordial and 'a' denotes astrophysical models --- for a source at a 
distance of $250\,\text{Mpc}$. The shaded band indicates the uncertainty due to the arbitrary 
spatial position of the sources. The dashed black line in the right figure indicates the upper limit of the trajectory 
length of $4423\,\text{Mpc}$ due to the Hubble time.}
\label{rmsDefl_hack}
\end{figure}

In contrast, the Fig.~\ref{rmsDefl_hack} shows that the EGMF models by H+18 predict significantly larger 
deflections and trajectory lengths, in particular for the primordial models due to a significantly higher initial seed 
field strength compared to D+05. Here, the CR candidates from Cygnus A can be expected to provide the necessary 
distribution of arrival directions in order to hold the conclusions from E+18. However, at rigidities below a few 
hundreds of PV $\bar{l}_{\rm traj}$ exceeds the upper limit that is given by the Hubble time. Thus, the primordial 
H+18 models also yield that Cygnus A is beyond the magnetic horizon at $R\lesssim 1\,\text{EV}$. In the case of the 
astrophysical H+18 models \emph{astrophysicalR} and \emph{astrophysical1R}\footnote{The \emph{astrophysicalR} model 
assumes an energy budget per feedback episode of $10^{60}\,\text{erg}$ from $z=4$, whereas a changing budget with 
$(10^{60}\dots5\times10^{58})\,\text{erg}$ for $z=1\dots 0$ is supposed in the \emph{astrophysical1R} model.} the 
resulting CR deflection are significantly larger than in the case of the D+05 model, but still below 
$\bar{\theta}_{\text{iso}}$ at $\sim 8\,\text{EV}$. The large uncertainties at small rigidities for models with a high 
cumulative filling factor indicate that the chosen spatial position of the source has a significant impact on the 
outcome. Hence, the inverted simulation setup only provides accurate results at $R\gtrsim 0.5\,\text{EV}$ for these 
models.

Summing up, a few single, individual sources, like Cygnus A and Centaurus A, will not be the only dominant HECR sources 
above the ankle if the EGMF provides a cumulative filling factor of about the astrophysical H+18 model or below --- as 
in the case of the D+05 EGMF. 
In this case, the dominant contribution up to $\sim 8\,\text{EeV}$ needs to be provided by a 
multitude of homogeneously and isotropically distributed sources, as pursued in the following.
\section{FR radio galaxies as HECR emitters}
\label{Sec:nonlocalRGs} 
The E+18 model already showed that the average non-local source population according to the local radio luminosity 
function (RLF) from Mauch and Sadler \cite{2007MNRAS.375..931M} cannot explain the observed spectral behavior above the 
ankle. But this RLF does not cover the contribution of (rather distant) high-luminous FR-II sources, that dominates 
the RLF at $L_{151}\gtrsim 10^{26.5}\,\text{W}\,\text{Hz}^{-1}\,\text{sr}^{-1}$ in the non-local Universe 
\cite{doi:10.1046_j.1365-8711.2001.04101.x}. In addition, the kinetic power of the jet most likely also depends on the 
FR classification of the source due to different lobe dynamics \cite{doi:10.1093_mnras_stv2712}. In order to 
constrain the HECR contribution by the bulk of the different types of FR RGs, an appropriate continuous source function 
(CSF) is needed. 

Therefore, the calculations from E+18 are repeated using the RLF from Willott et 
al.~\cite{doi:10.1046_j.1365-8711.2001.04101.x} --- hereafter referred as W+01 --- that differentiates between the FR 
types and includes the redshift dependence according to the source evolution. 
In addition, the impact of different ratios of radio luminosity $L_{\rm radio}$ to jet power $Q_{\rm jet}$, also known 
as the radiative efficiencies, are investigated. Due to the lack of reliable empirical methods to measure the jet power 
\cite{doi:10.1093_mnras_stv2712}, there are plenty of studies on this issue providing slightly different results. 
Willott et al.~\cite{1999MNRAS.309.1017W} --- hereafter referred as W+99 --- have derived a popular, model 
dependent predictor of the jet power of FR-II sources implying a systematic uncertainty $f^{3/2}$ with $1\leq f \leq 
20$. Other analysis have confirmed this $L_{\rm radio}$\,--\,$Q_{\rm jet}$ correlation even for FR-I sources 
\cite{2004MNRAS_349_1419C} within the uncertainty band. However, most of the other predictions yield a rather high $f$ 
value \cite{2000AJ_119_1111B} and a slightly different slope of the correlation \cite{B_rzan_2008, Cavagnolo_2010}. 
Godfrey and Shabala \cite{0004-637X-767-1-12, doi:10.1093_mnras_stv2712} --- hereafter referred as GS13 and GS16, 
respectively --- investigated the hypothesis of a significant difference in the distribution of the energy budget 
between FR-I and FR-II sources that has not been taken into account so far: In FR-I RGs the energy budget is dominated 
by a factor of $\gg 100$ by non-radiating particles yielding a rather high $f$ value, while radiating particles dominate 
this budget in the lobes of FR-II RGs suggesting a low $f$ value. However, the expected difference in the normalization 
of the $L_{\rm radio}$\,--\,$Q_{\rm jet}$ correlation is not observed, and also the theoretically expected difference 
in the slope $\beta_L$ of the correlation, due to different jet dynamics, could not be verified so far. 

The radio-to-CR correlation provides the energy density in CRs as 
\be
Q_{\rm cr}=\frac{g_{\rm m}}{1+k}\,Q_{\rm jet}=\frac{g_{\rm m}}{1+k}\,Q_0\,\left( \frac{L_{151}}{L_{p}} 
\right)^{\beta_L}
\label{Qcr}
\ee
where $g_{\rm m}$ denotes the fraction of jet energy found in leptonic and hadronic matter and the ratio of leptonic to 
hadronic energy density is given by $k$. 
Here, all of the introduced parameters differentiate between FR-I and FR-II. 
In principle, $g_{\rm m}<1$ and in the case of a minimum-energy magnetic field this parameter yields $g_{\rm m}\simeq 
4/7$ \cite{1970ranp.book.....P}. Note, that deviations from the given correlation (\ref{Qcr}) at the order of more than 
a magnitude occur for individual sources. Based on the most recent models by Godfrey and Shabala the normalization 
$Q_0$ is estimated by equalizing the jet power at the pivot luminosity
\be
L_{p} = \begin{cases}
              10^{24}/(4\pi)\,\text{W}\,\text{Hz}^{-1}\,\text{sr}^{-1} &\quad\text{ for FR-I at 151\,MHz}\,, \\
              10^{27.6}/(4\pi)\,\text{W}\,\text{Hz}^{-1}\,\text{sr}^{-1} &\quad\text{ for FR-II at 151\,MHz}\,,
             \end{cases}
\ee
taken from GS16, to the corresponding jet power given by the GS13 model, which yields
\be
Q_0 \simeq \begin{cases}
              2.27\times 10^{43}\,\text{erg/s} &\quad\text{ for FR-I}\,, \\
              3.04\times 10^{45}\,\text{erg/s} &\quad\text{ for FR-II}\,.
             \end{cases}
\label{eq:normalization}
\ee
Here, a rather large normalization factor ($g=2$) for the GS13 correlation model of FR-II RGs is supposed. 
GS16 showed that the empirical methods are strongly affected by the distance dependence, and basically 
the whole range of $0.5\lesssim\beta_L\lesssim 1.4$ is possible \cite{doi:10.1093_mnras_stv2712}. Therefore the 
authors suggest a theoretical approach which leads to a slope of 
\be
\beta_L = \frac{2}{(3+\alpha)(1-n_Q/n_t)}
\ee
if the lobe dynamics are parameterized by $V\propto t^{n_t}\,Q_{\rm jet}^{n_Q}$, where $V$ denotes the lobe volume at a 
given time $t$. Using a typical radio spectral index\footnote{The flux density $S_\nu$ at frequency 
$\nu$ is determined by the radio spectral index according to $S_\nu\propto \nu^{-\alpha}$.} $\alpha\simeq 0.8$, as 
well as the supersonic, self-similar lobe model \cite{1999MNRAS.309.1017W} for FR-II RGs and the buoyancy lobe model 
\cite{2004ApJ...607..800B, Cavagnolo_2010} for FR-I RGs, Godfrey and Shabala obtain
\be
\beta_L \simeq \begin{cases}
              0.5 &\quad\text{ for FR-I}\,, \\
              0.8 &\quad\text{ for FR-II}\,.
             \end{cases}\\
\ee
Note, that in the case of powerful FR-I RGs a steeper slope in the range $0.5\lesssim \beta_L \lesssim 0.8$ is expected. 

\begin{figure}[tbh]
  \centering
    \includegraphics[width=0.69\textwidth]{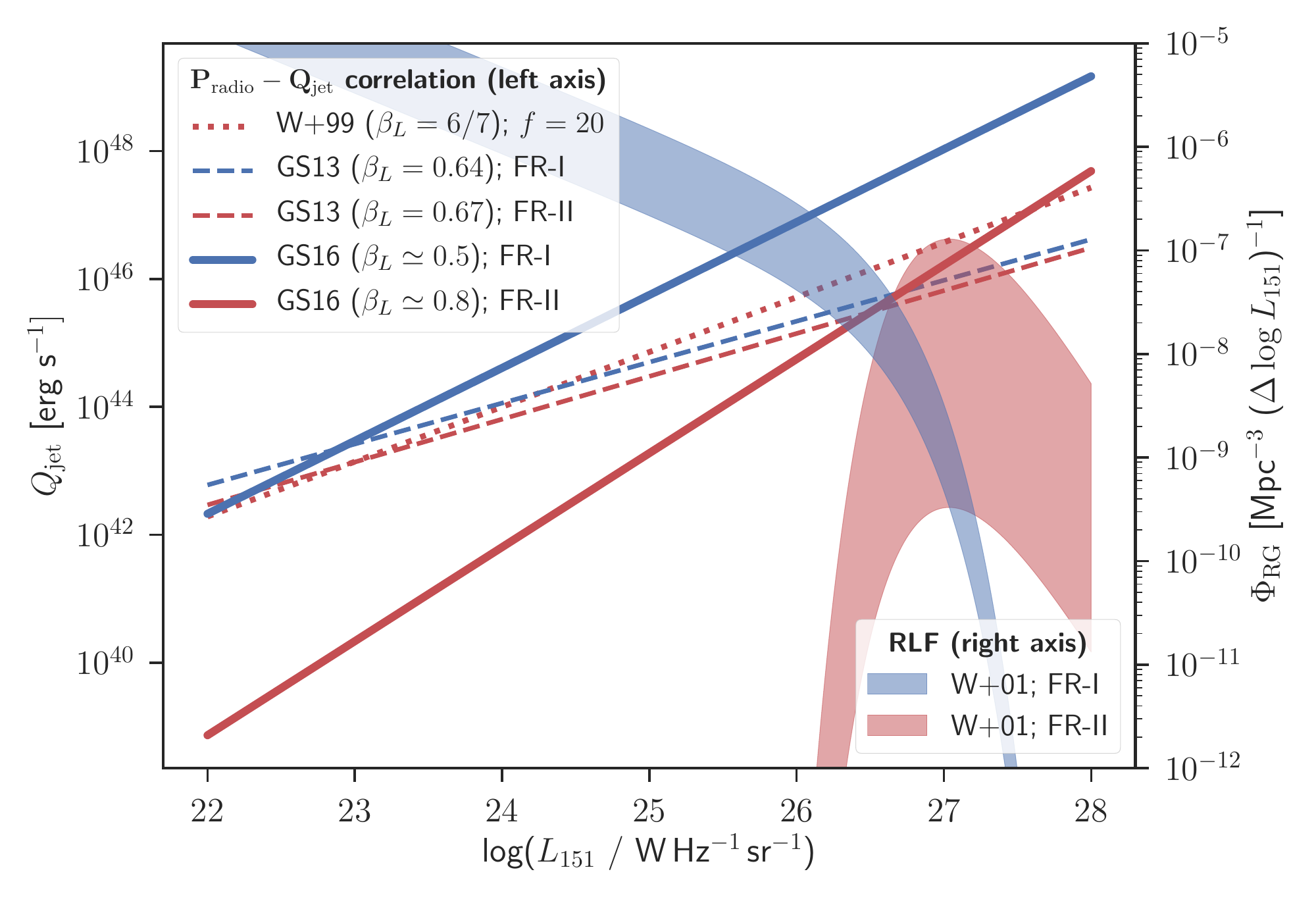}
\caption{Different models of the radio to jet power correlation (left axis) and the RLF of W+01 derived for model A for 
an open cosmology ($\Omega_{\rm M}=0$) with a redshift $z\in [0,\,2]$.}
\label{jet2radio}
\end{figure}

The Fig.~\ref{jet2radio} shows that the W+99 model 
is in good agreement with the FR-II prediction by the GS16 model in the case of low $f$ values. Taking the upper limit 
of $f$ seriously, the normalization (\ref{eq:normalization}) cannot exceed $10^{46}\,\text{erg/s}$ for FR-II. As 
expected from theory, the predicted jet power of low-luminous FR-I RGs by Godfrey and Shabala is above the W+99 
prediction, and the flat slope of the correlation yields a significant increase of the CR contribution by low-luminous 
FR-I.

In the large scale structures $\gtrsim 1\,\text{pc}$ of radio galaxies the dominant loss time scale is given by the 
escape time \cite{1475-7516-2018-02-036} $\tau_{\rm esc} \simeq r/(\beta_{\rm sh}c)$ which is estimated by the shock 
or shear velocity $\beta_{\rm sh}c$ and the size $r$ of the jet. For the common assumption of Bohm 
diffusion, which applies at non-relativistic shocks \cite{2006NatPh...2..614S, 2007Natur.449..576U}, the 
acceleration takes place on a timescale 
$\tau_{\rm acc}=f_{\rm diff}\,r_{\rm L}/(c \beta_{\rm sh}^2)$
for cosmic-ray particles with a Larmor radius $r_{\rm L}=R/B$. Here $R$ denotes the particle rigidity, and 
$1\lesssim f_{\rm diff}\lesssim 8$ encapsulates all details of the upstream and downstream plasma properties 
\cite{doi:10.1093/mnras/stt179} in a strongly turbulent magnetic field for standard geometries \cite{Drury:1983zz}. 
In steady state, the equality of both time scales yields the maximal rigidity
\be
\hat{R}\equiv {E_{\rm max} \over Z\,e} = {\beta_{\rm sh} \over f_{\rm diff}}\,B\, r = g_{\rm acc} \sqrt{\frac{(1-g_{\rm 
m})Q_{\rm jet}}{c}}\,,
\label{Rmax}
\ee
where the magnetic field power of the jet $Q_B=c\,\beta_{\rm jet}\,\pi 
r^2\,B^2/8\pi = Q_{\rm jet}(1-g_{\rm m})$ is used. Here, the acceleration efficiency parameter
\be
g_{\rm acc}=\sqrt{\frac{8\,\beta_{\rm sh}^2}{f_{\rm diff}^2\,\beta_{\rm jet}}}\;
\ee
is introduced and in the case of the typical shock and jet velocities $\beta_{\rm sh} \sim \beta_{\rm jet}\sim 0.1$ in 
extended jets of radio galaxies yielding $0.01 \lesssim g_{\rm acc} \lesssim 1$.
Note, that only non-relativistic shocks are considered here, since relativistic ones are poor accelerators 
to EeV energies \cite{10.1093/mnras/stu088, 10.1093/mnras/stx2485}. However, mildly relativistic, parallel shocks 
($0.2\lesssim \beta_{\rm sh} \lesssim 0.5$) are expected to be good UHECR accelerators \cite{10.1093/mnras/stx2485}, 
which still leads to $g_{\rm acc}\sim 1$ at most. So, the suggested range of $g_{\rm acc}$ also includes the case of a 
mean free path $\lambda\sim r_L^2/s$, where the magnetic field is randomly orientated on a scale-size $s<r$, or the case 
of discrete, mildly relativistic shear acceleration \cite{PhysRevD.97.023026}. In addition, the shear acceleration 
scenario provides a hard initial CR spectrum with a spectral index $a\leq 1$, that yields a high energy budget in 
UHECRs. 

Unless otherwise stated, the typical parameter values $g_{\rm acc}=0.1$ and $g_{\rm m}=4/7$ are used in the following. 
\subsection{Continuous Source Function}
\label{Sec:CSF} 
The number of radio sources per volume per power bin yields
\be
\frac{\diff N}{\diff V\,\diff Q_{\rm cr}} = \frac{\Phi_{\rm 
RG}(L_{151},\,z)}{2.3\,\beta_L\,Q_{\rm 
cr}}\,,
\ee
where $\Phi_{\rm RG}$ denotes the RLF from W+01 of (i) low-luminous radio sources, including FR-I as 
well as FR-II sources with low-excited/weak emission lines, and (ii) high-luminous radio sources, composed almost 
exclusively of sources with FR-II radio structures, respectively. 
In the following, the differentiation of $\Phi_{\rm RG}$ based on the FR type is simplified using 
\be
\frac{\diff N}{\diff V\,\diff Q_{\rm cr}} = 
 \begin{cases}
 \frac{\rho_{{\rm l} \circ}}{2.3\,\beta_L\,Q_{\rm cr}}\, \left( \frac{Q_{\rm cr}}{g_{\rm m}\,Q_\star}  
\right)^{-\alpha_{\rm l}/\beta_L} \, \exp\left(- \left( \frac{Q_{\rm cr}}{g_{\rm m}\,Q_\star} 
\right)^{1/\beta_L}  \right) 
\, f_I(z) \,,&\quad\text{for FR-I}\,,\\
 \frac{\rho_{{\rm h} \circ}}{2.3\,\beta_L\,Q_{\rm cr}}\, \left( \frac{Q_{\rm cr}}{g_{\rm m}\,Q_\star}  
\right)^{-\alpha_{\rm h}/\beta_L} \,\exp\left(- \left( \frac{g_{\rm m}\,Q_\star}{Q_{\rm cr}}
\right)^{1/\beta_L}  \right) 
\, f_{II}(z) \,,&\quad\text{for FR-II}\,,
\end{cases}
\ee
where
\be
\begin{split}
Q_\star &= \begin{cases}
   \left( \frac{4\,\pi L_{{\rm l} \star}}{L_p} \right)^{\beta_L} \, Q_0\,,&\quad\text{for FR-I}\,,\\
   \left( \frac{4\,\pi L_{{\rm h} \star}}{L_p} \right)^{\beta_L} \, Q_0\,,&\quad\text{for FR-II}\,,\\
               \end{cases}\\
f_I(z) &=\begin{cases}
          (1+z)^{k_{\rm l}} &\quad\text{for }\,z<z_{{\rm l}\circ}\,,\\
          (1+z_{{\rm l}\circ})^{k_{\rm l}} &\quad\text{for }\,z\geq z_{{\rm l}\circ}\,,
         \end{cases}\\
f_{II}(z) &=\begin{cases}
          \exp\left(-\frac{1}{2} \left( \frac{z-z_{{\rm h}\circ}}{z_{{\rm h}1}} \right)^2 \right) &\quad\text{for model 
A or models B and C at }\,z<z_{{\rm h}\circ}\,,\\
          1 &\quad\text{for model B at}\,z\geq z_{{\rm h}\circ}\,,\\
          \exp\left(-\frac{1}{2} \left( \frac{z-z_{{\rm h}\circ}}{z_{{\rm h}2}} \right)^2 \right) &\quad\text{for model 
C at 
}\,z\geq z_{{\rm h}\circ}\,,
         \end{cases}
\end{split}
\ee 
for two different cosmologies with $\Omega_{\rm M}=1$ and $\Omega_{\rm M}=0$ for three different parameter 
models A, B, C. The model dependent best-fit parameters from W+01 are given in Table \ref{tab:rlffit}. 

\begin{table*}
\footnotesize
\begin{center}
\begin{tabular}{ccccccccccccc}
\hline\hline 
\mc{1}{c}{Model} &\mc{1}{c}{$\Omega_{\rm M}$} &\mc{1}{c}{$\log(\rho_ {{\rm l}\circ})$}&\mc{1}{c}{$\alpha_{\rm 
l}$}&\mc{1}{c}{$\log(L_{{\rm l} \star})$}&\mc{1}{c}{$z_{{\rm l}\circ}$}&\mc{1}{c}{$k_{\rm l}$}&\mc{1}{c}{$\log$ ($\rho_ 
{{\rm h}\circ}$)}&\mc{1}{c}{$\alpha_{\rm h}$}&\mc{1}{c}{$\log(L_{{\rm h} \star})$}&\mc{1}{c}{$z_{{\rm 
h}\circ}$}&\mc{1}{c}{$z_{{\rm h}1}$}&\mc{1}{c}{$z_{{\rm h}2}$}\\

\hline\hline

A & 1 & $-7.153$ & $0.542$ & $26.12$ & $0.720$ & $4.56$ 
  & $-6.169$ & $2.30$  & $27.01$ & $2.25$  & $0.673$ & -- \\

B & 1 & $-7.150$ & $0.542$ & $26.14$ & $0.646$ & $4.10$
  & $-6.260$ & $2.31$  & $26.98$ & $1.81$  & $0.523$ & -- \\

C & 1 & $-7.120$ & $0.539$ & $26.10$ & $0.706$ & $4.30$ 
  & $-6.196$ & $2.27$  & $26.95$ & $1.91$  & $0.559$ & $1.378$ \\
\hline
A & 0 & $-7.503$ & $0.584$ & $26.46$ & $0.710$ & $3.60$ 
  & $-6.740$ & $2.42$  & $27.42$ & $2.23$  & $0.642$ & -- \\

B & 0 & $-7.484$ & $0.581$ & $26.47$ & $0.580$ & $3.11$ 
  & $-6.816$ & $2.40$  & $27.36$ & $1.77$  & $0.483$ & -- \\

C & 0 & $-7.523$ & $0.586$ & $26.48$ & $0.710$ & $3.48$
  & $-6.757$ & $2.42$  & $27.39$ & $2.03$  & $0.568$ & $0.956$ \\

\hline\hline         
\end{tabular}
\end{center} 
{\caption[Table of observations]{\label{tab:rlffit}Best-fit parameters
for RLF models A, B and C for $\Omega_{\rm M}=1$ and $\Omega_{\rm M}=0$, respectively, taken from W+01. Here, $\rho_ 
{{\rm 
l}\circ}$, $\rho_ {{\rm h}\circ}$ are in units of $\text{Mpc}^{-3}\,(\Delta \log(L_{151}))^{-1}$ and $L_{{\rm 
l} \star}$, $L_{{\rm h} \star}$ are in units of $\text{W}\,\text{Hz}^{-1}\,\text{sr}^{-1}$.}}
\normalsize
\end{table*}

Thus, the redshift dependent CSF of FR-I and FR-II sources, respectively, is given by
\begin{equation}
\Psi_{i}(R,\,z) \equiv {\mathrm{d}N_{\rm cr}(Z_i) \over \mathrm{d}V \mathrm{d}R\,\mathrm{d}t} = 
\int_{\check 
Q_{\rm cr}}^{\hat Q_{\rm cr}} S_i\big(R,\hat R(Q_{\rm cr})\big)\,{\mathrm{d}N \over 
\mathrm{d}V\,\mathrm{d}Q_{\rm cr}}\,\mathrm{d}Q_{\rm cr} 
\label{CRsourceRateDensity}
\end{equation}
where \mbox{$S_i(R,\hat R(Q_{\rm cr})) \equiv \diff N_{\rm cr}(Z_i)/\diff R\,\diff t$} denotes the cosmic ray 
spectrum of element species $i$ with charge number $Z_i$, emitted by a FR-I/II source with total cosmic ray power per 
charge number, $Q_{{\rm cr},i}\equiv Q_{\rm cr}(Z_i)=f_i\,Z_i\,Q_{\rm cr}/\bar Z$, up to a maximal rigidity $\hat 
R(Q_{\rm cr})$. The limits of integration are the smallest, $\check Q_{\rm cr}$, respectively largest, $\hat Q_{\rm 
cr}$, CR powers that need to be considered.

To solve this integral analytically, one has to suppose that the individual source spectra are given by 
\begin{equation}\label{eq:siglespec}
S_i(R,\hat R(Q_{\rm cr})) = \nu_i(a)\,Q_{\rm cr}\,\left({R \over \check R}\right)^{-a}\,\Theta\big(\hat R(Q_{\rm cr}) - 
R\big)\;,
\end{equation}
with the Heaviside step function $\Theta(x)$ that introduces a sharp cutoff at 
\begin{equation}
\hat R(Q_{\rm cr}) = g_{\rm acc} \sqrt{\frac{1}{c}\,\left(\frac{1}{g_{\rm m}}-1\right)\,(1+k)\,Q_{\rm cr}}
\label{maxRig}
\end{equation}
according to Eq.~(\ref{Rmax}). 
Analogous to the approach by E+18, the requirement 
\be
Q_{\text{cr},i}={f_i\,Z_i\,Q_{\rm cr} \over \bar Z} =eZ_i\int_{\check{R}}^{\hat{R}(Q_{\rm cr})} 
\text{d}R\,R\;S_i(R,\hat 
R(Q_{\rm cr}))\,,
\ee 
yields the spectral normalization correction $\nu_i(a)$ as
\be
\nu_i(a) = {f_i \over e \bar Z \check{R}^2}\times
\begin{cases}
\;(2-a)\,\big/\,(\rho_{\rm cr}^{2-a}-1)\;,& \text{ for }\, a\neq 2\,\\
\;1/\ln \rho_{\rm cr} \phantom{\Big|}\;,& \text{ for }\, a=2
\label{nonlocalNorm0}
\end{cases}
\ee
with the cosmic ray dynamical range $\rho_{\rm cr} \equiv \hat{R}(Q_{\rm cr})/\check{R}$. 
For a maximal CR power 
\begin{equation}
\hat Q_{\rm cr} > \frac{g_{\rm m}}{1+k}\,Q_{\star}\,\left( \frac{R}{R_{\star}} \right)^2
\end{equation}
the approximate analytical solution to Eq.\ (\ref{CRsourceRateDensity}) is given by
\begin{equation}
\Psi_{i}(R,z) \;\simeq\;
\begin{cases}
 \begin{split}
 & \frac{\rho_{{\rm l} \circ}\,f_i\,\nu_a\,c}{2.3\,e\,\bar 
Z}\,\left[g_{\rm acc}^2 \left(\frac{1}{g_{\rm m}}-1 \right) 
(1+k)\right]^{-1}\,\left(\frac{R}{R_\star}\right)^{-a}\,\frac{f_I(z)}{z+1}\, \\ & \times
\left[ \Gamma \left(\xi_{a}^{I},\, \left(\frac{R}{R_\star}\right)^{2/\beta_L}\right) - 
\Gamma\left(\xi_{a}^{I},\, \left(\frac{\hat Q_{\rm cr}(k+1)}{g_{\rm 
m}\,Q_{\star}}\right)^{1/\beta_L}\right) \right] \,,&\text{ for FR-I}\,,\\
& \frac{\rho_{{\rm h} \circ}\,f_i\,\nu_a\,c}{2.3\,e\,\bar 
Z}\,\left[g_{\rm acc}^2 \left(\frac{1}{g_{\rm m}}-1 \right) 
(1+k)\right]^{-1}\,\left(\frac{R}{R_\star}\right)^{-a}\,\frac{f_{II}(z)}{z+1}\, \\ & \times
\left[ \Gamma\left(\xi_{a}^{II},\, \left(\frac{g_{\rm m}\,Q_{\star}}{\hat Q_{\rm cr} (k+1)}\right)^{1/\beta_L}\right) - 
\Gamma \left(\xi_{a}^{II},\, \left(\frac{R_\star}{R}\right)^{2/\beta_L}\right)
\right] \,,&\text{ for FR-II}\,,
 \end{split}
\end{cases}
\label{ProdRateDens1}
\end{equation}
for the three simplifying cases
\begin{align*}
\nu_a &= 2-a\;; & \xi_{a}^{I} &= -\alpha_{\rm l}+a\beta_L/2\;; & \xi_{a}^{II} &= \alpha_{\rm h}-a\beta_L/2\;; 
&\qquad\text{for} &\;a<2,\,a\not\approx 2\\
\nu_a &= 1/\ln \rho_\star\;; & \xi_{a}^{I} &= -\alpha_{\rm l}+\beta_L\;; & \xi_{a}^{II} &= 
\alpha_{\rm h}-\beta_L\;; &\qquad\text{for} &\;a\simeq 2\\
\nu_a &= (a-2)\rho_\star^{2-a}\;; & \xi_{a}^{I} &= -\alpha_{\rm l}+\beta_L\;; & \xi_{a}^{II} 
&= \alpha_{\rm h}-\beta_L\;; &\qquad\text{for} &\;a>2,\,a\not\approx 2\,.
\end{align*}
Here, the critical rigidity
\be
R_\star = g_{\rm acc}\sqrt{\frac{(1-g_{\rm m})\,Q_\star}{c}} 
= \begin{cases}
   8.2\times 10^{18}\,g_{\rm acc}(1-g_{\rm m})^{\frac{1}{2}}\,g_{\rm l}^{\beta_L} \, 
\text{V}\,,&\quad\text{for 
FR-I}\,,\\
   9.5\times 10^{19}\,g_{\rm acc}(1-g_{\rm m})^{\frac{1}{2}}\,g_{\rm h}^{\beta_L} \, 
\text{V}\,,&\quad\text{for 
FR-II}\,,\\
               \end{cases}
 \label{critRig}
\ee
is introduced with the RLF model dependent parameters 
\be
\begin{split}
 g_{\rm l} & = \sqrt{\frac{4\pi\,L_{{\rm l}\star}}{L_p}}\simeq 39.77\dots61.6\\
 g_{\rm h} & = \sqrt{\frac{4\pi\,L_{{\rm h}\star}}{L_p}}\simeq 1.68\dots2.88
\end{split}
\ee
as well as the corresponding dynamic range $\rho_\star = R_\star/\check{R}$. 
\begin{figure}[tbh]
  \centering
    \includegraphics[width=0.60\textwidth]{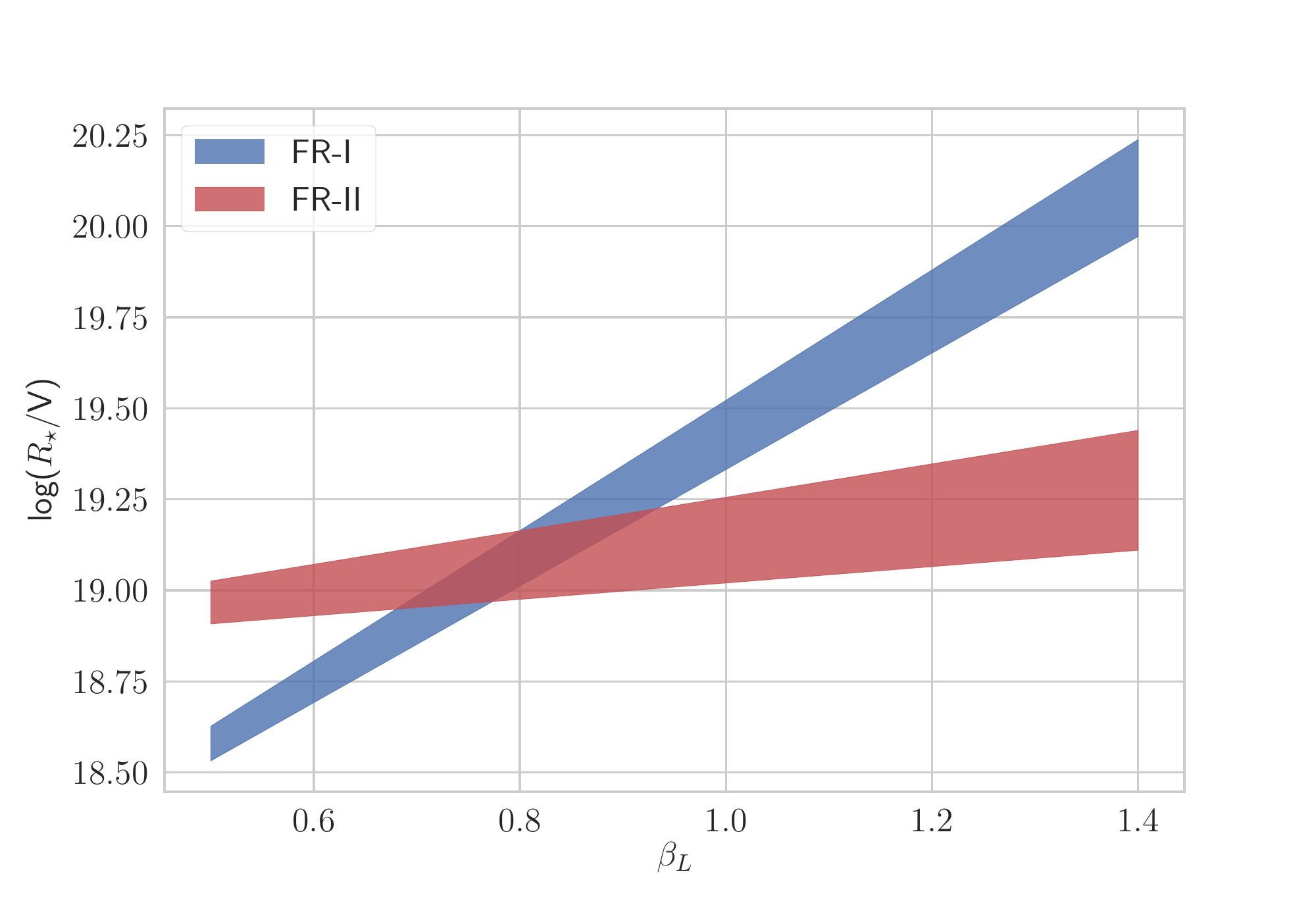}
\caption{The range of the critical rigidity dependent on $\beta_L$ for the different RLF models. Here and in the 
following the typical parameter values of $g_{\rm acc}=0.1$, $g_{\rm m}=4/7$ are used unless otherwise stated.}
\label{Rstar_betaL}
\end{figure}
\begin{figure}[tbh]
  \centering
    \includegraphics[width=0.49\textwidth]{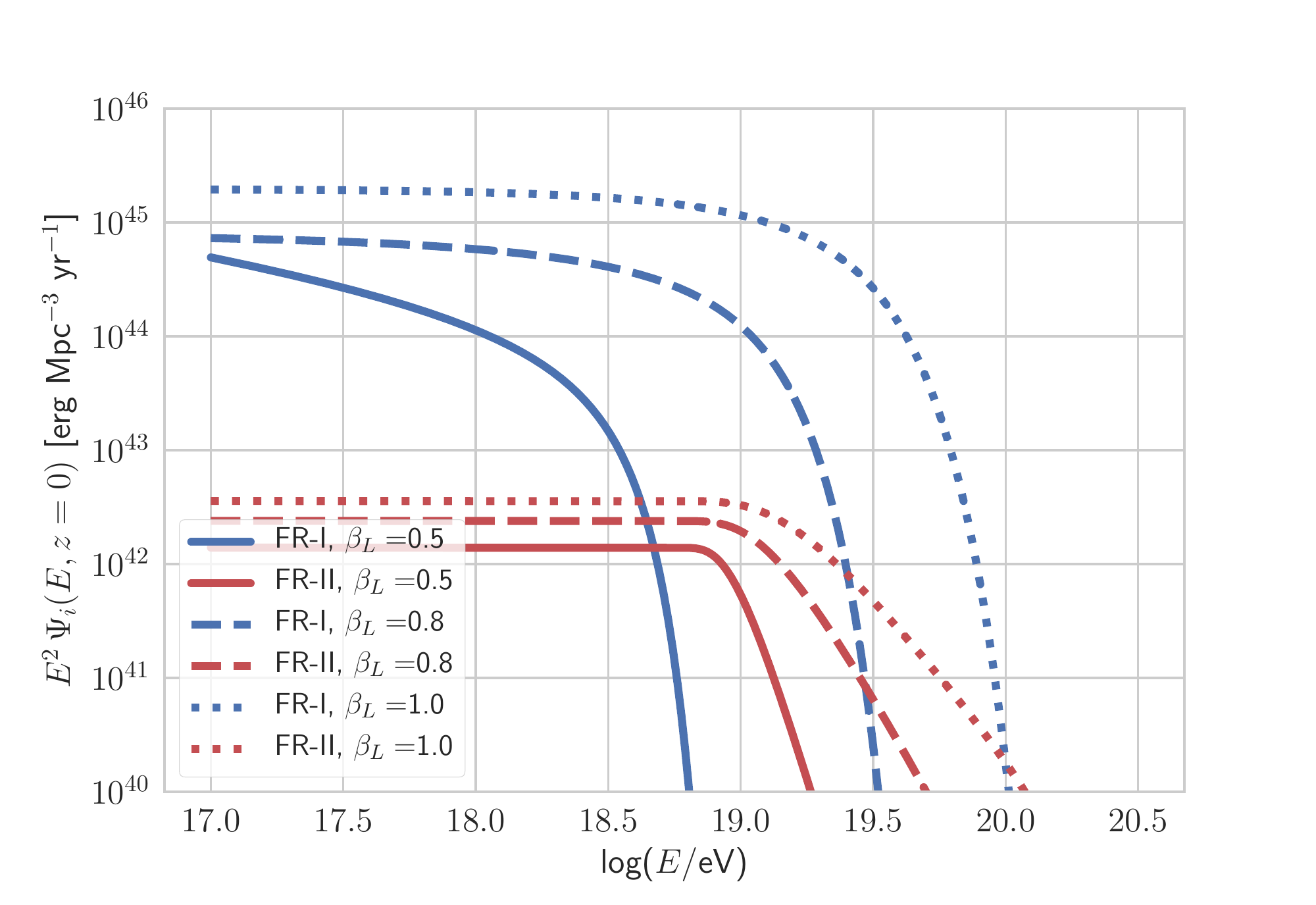}
    \includegraphics[width=0.49\textwidth]{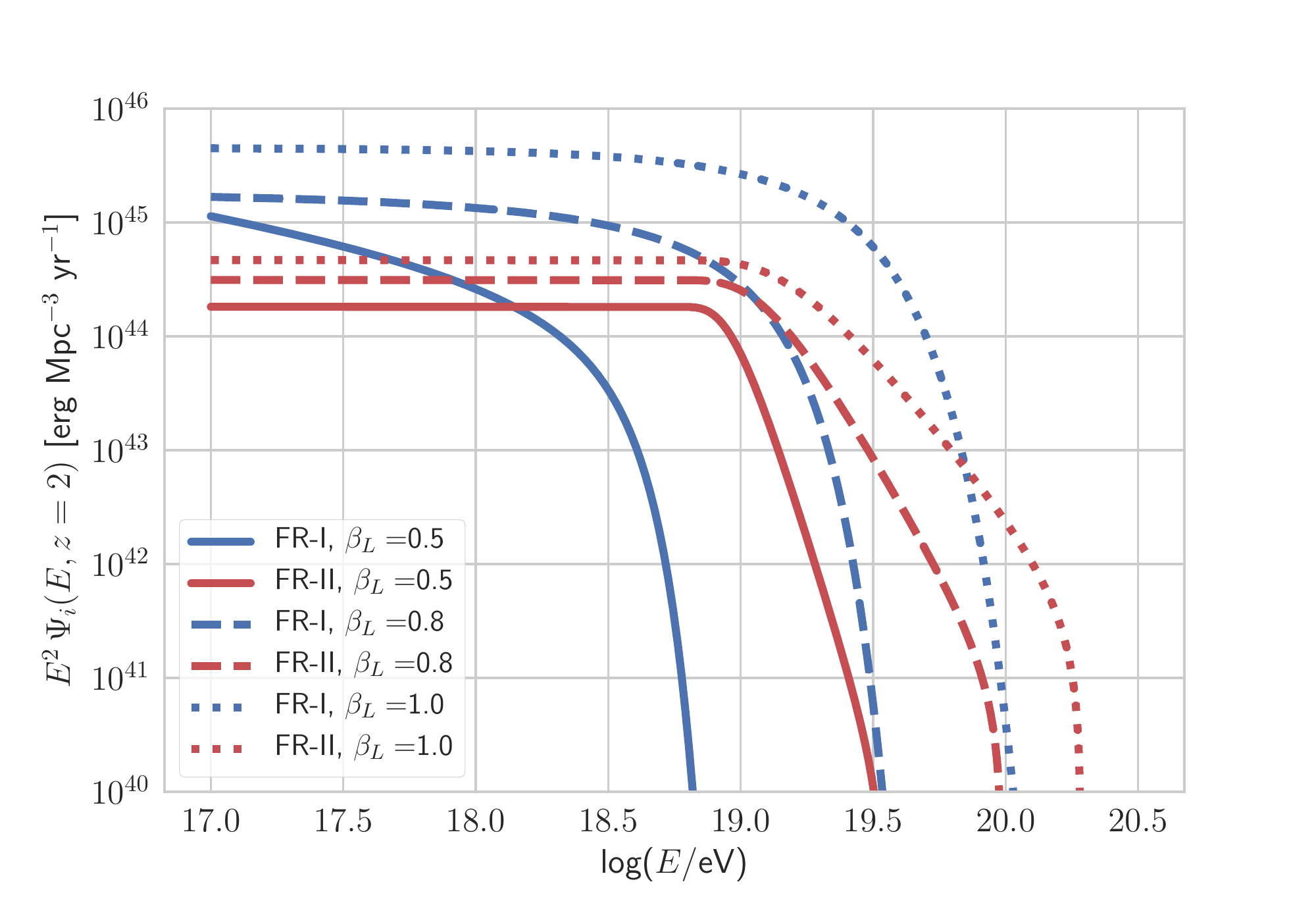}
\caption{CSF of CR protons from FR-I and FR-II sources with an initial spectral index $a=2$ and a vanishing leptonic 
energy fraction $k=0$ for different $\beta_L$ in 
the case of $z=0$ (\emph{left}) and $z=2$ (\emph{right}).}
\label{CSF_plot}
\end{figure}

The Fig.~\ref{Rstar_betaL} shows that the critical rigidity $R_\star$ of FR-I sources strongly depends on 
$\beta_L$, and in the case of the GS16 model 
\be
R_\star \sim \begin{cases}
   10^{18} \, \text{V}\,,&\quad\text{for FR-I if }\beta_L\simeq 0.5\,,\\
   10^{19} \, \text{V}\,,&\quad\text{for FR-II if }\beta_L\simeq 0.8\,
               \end{cases}
\ee
for the typical parameter values. 

Note, that the critical rigidity is a characteristic of the given distribution of RGs that results from the 
RLF model, so that it needs to be differentiated from the maximal rigidity of individual sources given by 
Eq.~\ref{maxRig}. Analyzing the asymptotic spectral behavior of the CSF (\ref{ProdRateDens1}) of FR-I and FR-II sources one recognizes 
that 
\be
\begin{split}
\Psi_{i}(R\ll R_\star,z) &\propto \left(\frac{R}{R_\star}\right)^{-a}\,,\\
\Psi_{i}(R\gg R_\star,z) &\propto \begin{cases}
              \left(\frac{R}{R_\star}\right)^{-a+2\xi_{a}^{I}/\beta_L-2/\beta_L}\,\exp\left( -
\left(\frac{R}{R_\star}\right)^{2/\beta_L} \right) &\quad\text{ for FR-I}\,, \\
              \left(\frac{R}{R_\star}\right)^{-a-2\xi_{a}^{II}/\beta_L} &\quad\text{ for FR-II}\,,
             \end{cases}
\end{split}
\ee
so that $R_\star$ denotes a spectral break, where the spectral behavior is no longer governed by 
the individual sources but gets steepened due to the impact of the RLF.  
Thus, the spectral behavior of the CSF of FR-I sources is hardly able to explain the observed CR spectrum at 
$E \gg 1\,\text{EeV}$ for $\beta_L \simeq 0.5$ (see Fig.~\ref{CSF_plot}), but these sources provide the 
necessary UHECR luminosity density of about $10^{44}\,\text{erg}\,\text{yr}^{-1}\,\text{Mpc}^{-3}$ 
\cite{2018arXiv180401064N}. These results are in good agreement with the ones from the E+18 model, as well as the 
luminosity density estimate from Matthews et al. \cite{doi:10.1093_mnras_sty2936}\footnote{Note, that the authors 
used the FR-I based radio-to-CR correlation from Cavagnolo et al.\ \cite{Cavagnolo_2010} and the local radio luminosity 
function from Heckman and Best \cite{doi:10.1146/annurev-astro-081913-035722}}.

In contrast, the spectral behavior of the CSF of FR-II sources is in principle 
able to explain the data. However, its contribution at small redshifts is significantly smaller than the contribution 
by FR-Is at about $1\,\text{EeV}$, if similar parameters of $g_{\rm acc}$, $g_{\rm m}$ and $a$ are supposed --- which is 
not necessarily the case. Nevertheless, the possible parameter range hardly enables FR-II sources to provide the necessary 
UHECR luminosity density, regardless of their UHECR contribution in the non-local Universe, as the magnetic horizon 
effect \cite{globusetal2008} limits the potential contributors to distances of a few hundreds of Mpc. Further details 
on the resulting HECR contribution at Earth including the impact of propagation effects are discussed in the 
following.
\subsection{Constraints on the HECR contribution}
\label{Sec:constraints}
Propagation effects need to be included in order to give an accurate estimate of the average contribution of the bulk 
of FR sources between $z=0$ and $z=2$ to the observed HECR data. Therefore, a 1D simulation is performed, as 
already introduced by E+18, where the production rate density (\ref{ProdRateDens1}) is used to obtain an 
absolutely normalized CR flux from the bulk of FR sources.
In general, a solar-like initial composition is supposed, i.e. 92\% H, 7\% He, 0.23\textperthousand\ C, 
0.07\textperthousand\ N, 0.5\textperthousand\ O, 0.08\textperthousand\ Si and 0.03\textperthousand\ Fe in terms of 
number of particles at a given rigidity. 
The chosen RLF model parameters hardly change the FR-I contribution, however, the FR-II contribution varies almost by 
an order of magnitude. Unless otherwise stated, the RLF model A for $\Omega_{\rm M}=1$ is used in the following, as 
this 
setup provides the maximal HECR contribution.
\begin{figure}[tbh]
  \centering
    \includegraphics[width=0.49\textwidth]{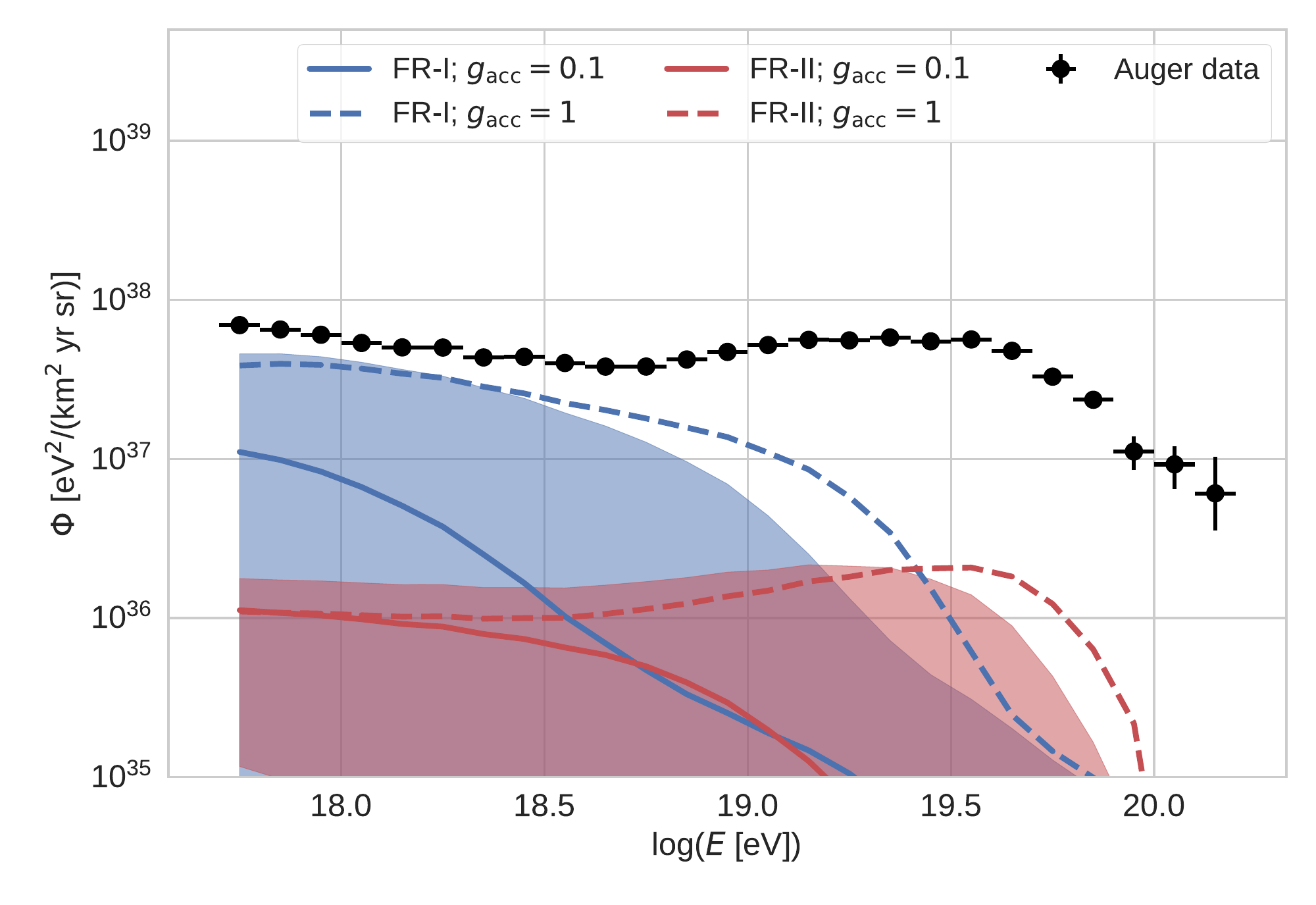}
     \includegraphics[width=0.49\textwidth]{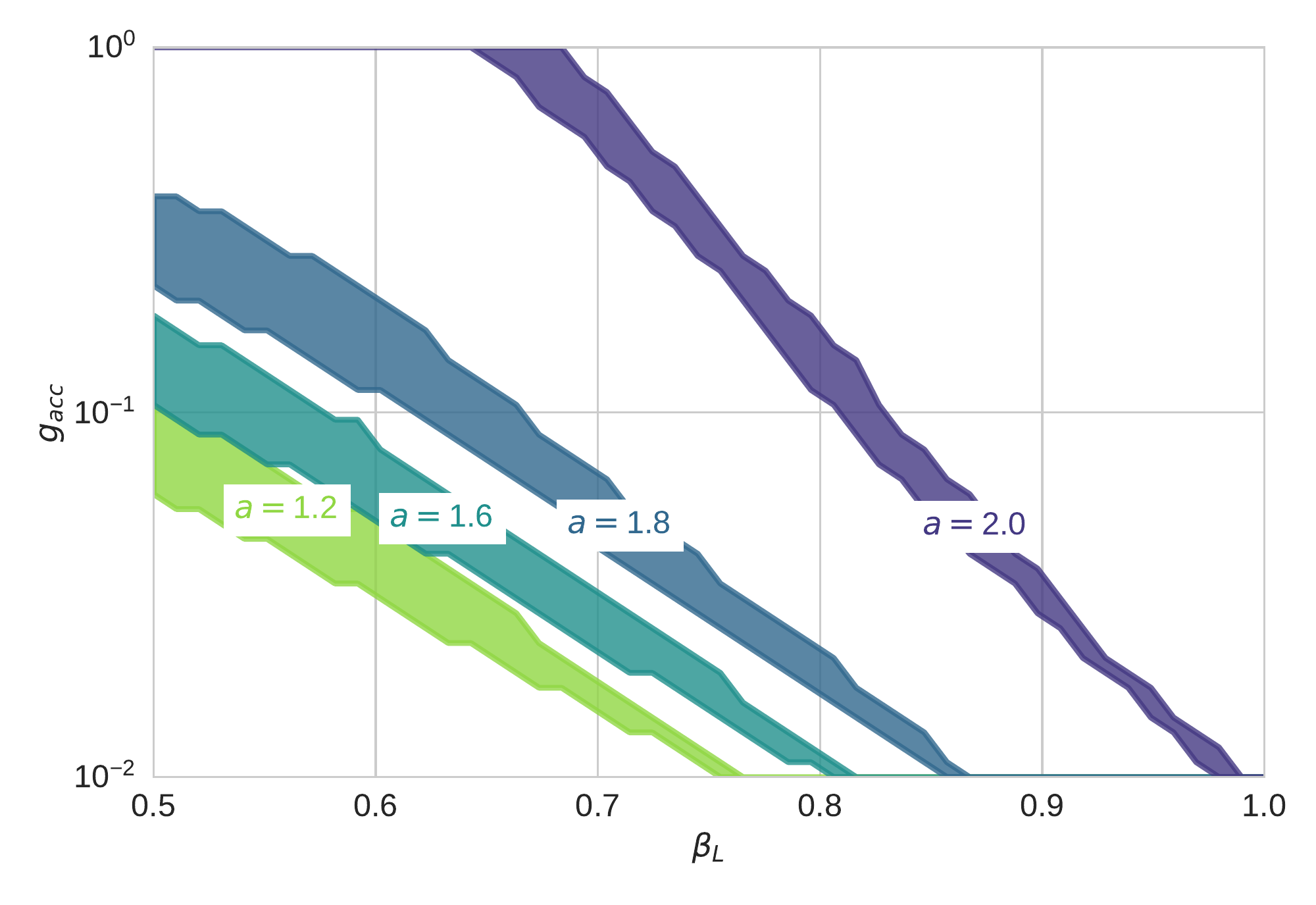}
\caption{The HECR contribution in the limiting case of $k=0$ and $g_{\rm m}=4/7$: \newline
\emph{Left:} HECR spectra by FR RGs using $a=2$ and the radio-to-CR correlation by GS16. The shaded areas indicate the 
results for $g_{\rm acc}\in [1,\,0.01]$ and $g_{\rm m}\in [0.9,\,0.1]$. \newline
\emph{Right:} The required acceleration efficiency $g_{\rm acc}$ of FR-I RGs dependent on $\beta_L$ for different 
spectral indexes $a$ of the initial CR spectrum. The shaded area indicates the uncertainty due to the different 
parametrization of the RLF of W+01.}
\label{GS16contrAndLim}
\end{figure}

In the case of the radio-to-CR correlations of GS16 (or GS13) the average HECR contribution by FR-II sources is even 
for a high acceleration efficiency and a high cosmic ray load at least a magnitude below the data, as shown in the left 
Fig.~\ref{GS16contrAndLim}, although, its spectral behavior looks quite promising, as already exposed several years ago 
\cite{Rachen:1992pg, Rachen:1993gf}. Further, it can be shown that even 
for a hard initial CR spectrum, i.e.\ $a\ll 2$, the FR-II contribution stays below the data points. Dependent on 
the critical rigidity $R_\star$, FR-I sources can provide the HECR flux below the ankle --- especially for 
$\beta_L\sim0.5$, as suggested by GS16 --- or above for sufficiently large $\beta_L$ and $g_{\rm acc}$ values. The 
right Fig.~\ref{GS16contrAndLim} explores the required parameter space of FR-I RGs 
in order to provide a significant contribution of HECRs. So, the typical first order Fermi 
acceleration spectrum will hardly result in a significant contribution by FR-I RGs, if the leptonic energy budget in 
the jets is not vanishing, i.e. $k\gtrsim 1$. But in the case of $a\ll2$, a significant contribution from these sources 
is expected, even for a small $\beta_L$ value if $g_{\rm acc}\gtrsim 0.1$.

\begin{figure}[tbh]
  \centering
  \includegraphics[width=0.49\textwidth]{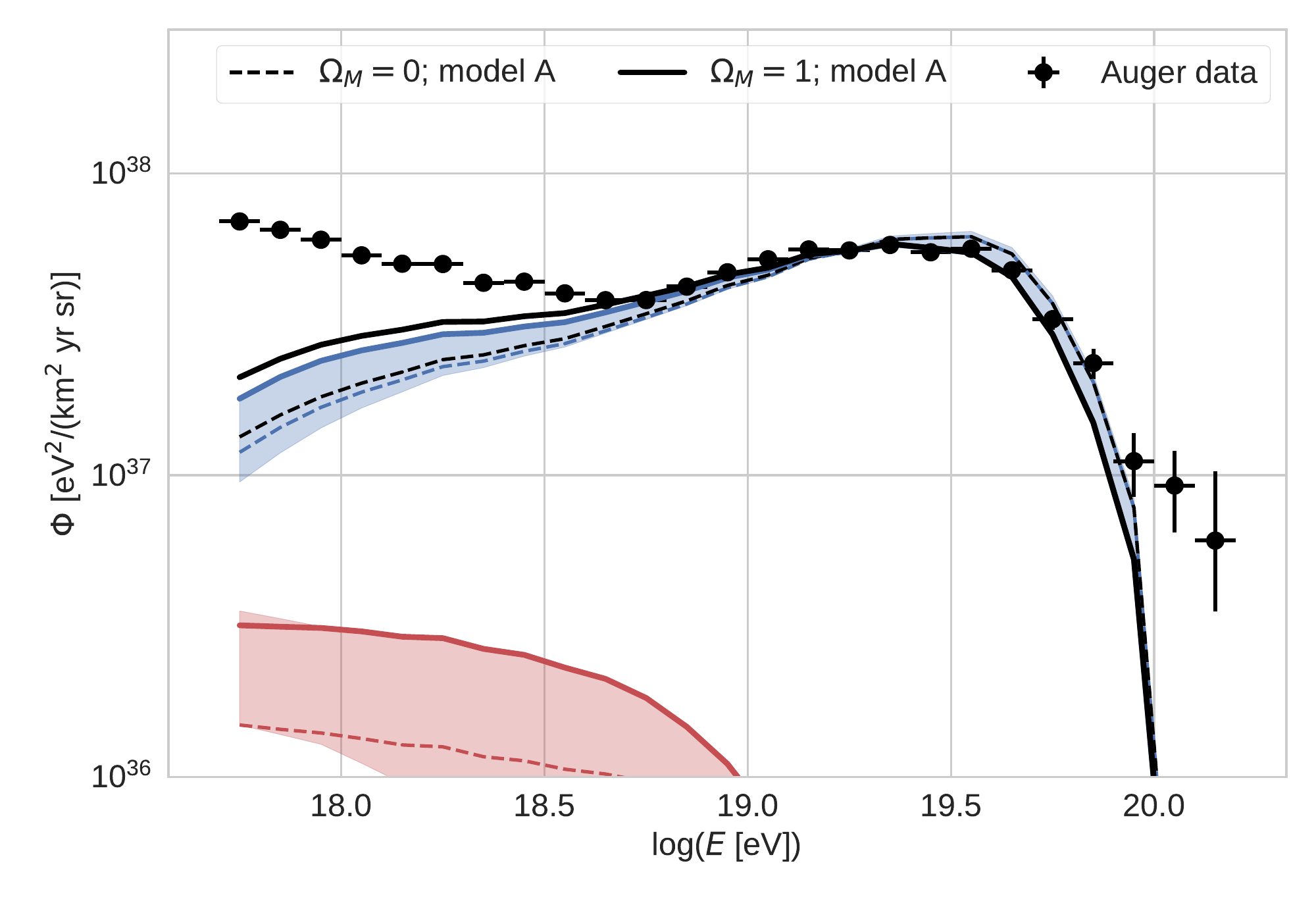}
  \includegraphics[width=0.49\textwidth]{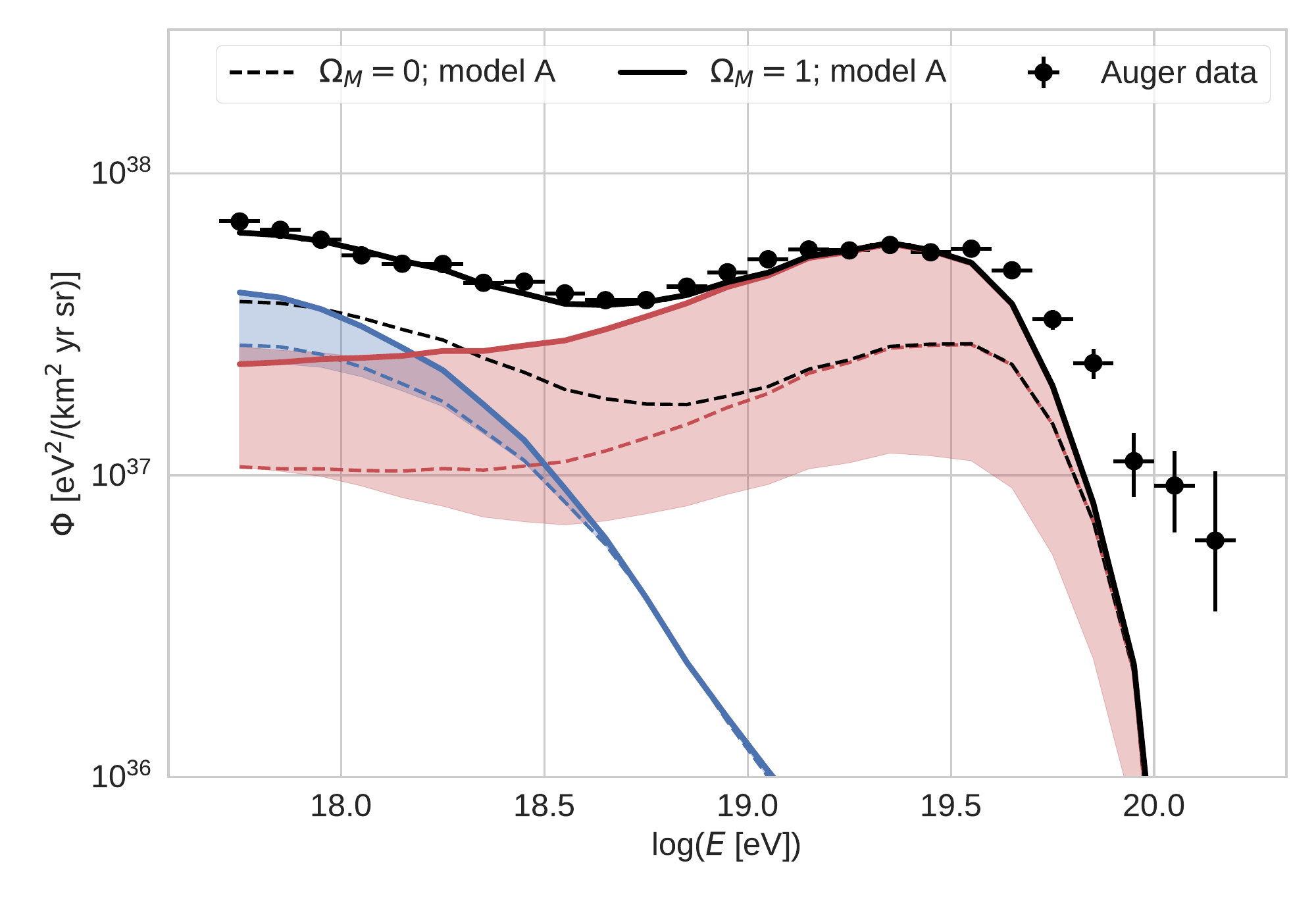}
\caption{Proof of principle fit scenarios, where the blue/ red lines indicate the individual contributions by FR-I/ 
FR-II RGs and the shaded bands expose the uncertainty due to the different RLF models: \newline
\emph{Left:} Scenario I with $a=1.8$, $g_{\rm m}=4/7$ for both FR classes; $\beta_L=0.9$, $k=12$, $g_{\rm acc}=0.8$ 
for FR-I RGs; and $\beta_L=0.8$, $k=0$, $g_{\rm acc}=0.1$ for FR-II RGs. \newline
\emph{Right:} Scenario II with $k=0$, $g_{\rm m}=4/7$, $g_{\rm acc}=0.2$ for both FR classes; $a=1.88$, $\beta_L=0.5$ 
for FR-I RGs; and $a=1.8$, $\beta_L=0.8$, as well as a modified normalisation $Q_0= 2.8\times 
10^{46}\,\text{erg\,s}^{-1}$ for FR-II RGs.}
\label{fitScenarios}
\end{figure}

Based on a simple trial-and-error fitting method, the Fig.~\ref{fitScenarios} introduces two scenarios that provide an 
accurate CR flux at $10^{18.7}\,\text{eV}\lesssim E \lesssim 10^{19.5}\,\text{eV}$ by FR-I RGs (scenario I) and 
FR-II RGs (scenario II), respectively. For the scenario I, a rather high $\beta_L$ value and a high acceleration 
efficiency are needed to obtain a critical rigidity (\ref{critRig}) above $\sim 10^{19.5}\,\text{V}$, so that the 
spectral behavior above the ankle becomes appropriate. For the scenario II, the jet power of FR-II RGs needs to exceed 
$10^{46}\,\text{erg\,s}^{-1}$ at the pivot luminosity as well as $k\sim 0$, $g_{\rm m}\sim 4/7$ and $a\lesssim 1.8$ in 
order to provide enough UHECRs. Due to the impact of the Greisen-Zatsepin-Kuzmin (GZK) effect 
\cite{1966PhRvL..16..748G, 1966JETPL...4...78Z} both scenarios fail at the highest energies. In contrast to scenario I, 
the scenario II also yields an appropriate HECR flux below the ankle due to the contribution by FR-I RGs. Note, that the 
necessary contribution from additional sources at higher and lower energies, respectively, most likely changes the given 
values of the fit parameters. 

Further, it has been checked that the resulting cosmogenic neutrino flux is even in the case of the scenario II below 
the current neutrino limits at energies above $0.1\,\text{EeV}$. 
Still the associated cosmogenic gamma-ray flux can be in tension with the isotropic diffusive gamma-ray background 
constraints by Fermi-LAT, due to the strong source evolution behavior of FR-II RGs \cite{2017ICRC...35..562V, 
Globus_2017}. However, the spectral index of the initial CR spectrum has about the same influence on the energy density 
of the diffusive gamma-ray background as the source evolution index \cite{2011PhRvD..84h5019A}. Thus, a rather hard CR 
spectrum with $a\ll 2$, as also suggested for a discrete, mildly relativistic shear acceleration scenario 
\cite{PhysRevD.97.023026}, is favored with the additional benefit of a higher HECR energy budget. 
However, more detailed fitting scenarios --- that include the inevitable contribution of a single (or multiple) 
individual, close-by source(s), as well as the other observational constraints of HECRs --- are needed, but beyond the 
scope of this work. 
\section{Conclusions}
\label{Sec:Conclusions}
In this paper, an extended D+05 EGMF structure and an efficient simulation setup are 
developed in order to examine the mean deflection $\bar{\theta}$ of CRs from distant sources. In the case of Cygnus A, 
CRs with a rigidity $\gtrsim 3\,\text{EV}$ yield $\bar{\theta}\lesssim 10\degree$, so that this source cannot provide 
the bulk of light CRs at around the ankle, where the observational data features no significant anisotropy so far. This 
leaves two possible conclusions:
\begin{enumerate}
 \item[(i)] The EGMF strength needs to be significantly higher than the one given by the D+05 model. Here, 
the H+18 models are probed as well, showing a substantially different outcome: Due to a significantly higher 
field strength in the large-scale structures of voids, filaments and sheets, all of the three primordial H+18 models 
yield UHECR deflections in the necessary order of magnitude for Cygnus A. However, at rigidities $\lesssim 
1\,\text{EV}$ the source is already beyond the magnetic horizon. 
 \item[(ii)] Cygnus A does not contribute significantly to the UHECR data, but a multitude of isotropically 
distributed sources, most likely radio galaxies or starburst galaxies \cite{Aab:2018chp}. Although, the latter source 
class might struggle to accelerate a nucleus up to the required rigidities \cite{doi:10.1093_mnrasl_sly099}.
\end{enumerate}
Based on the common radio to jet power correlations, this work determines the average HECR contribution of 
the different types of FR RGs dependent on the CR load of the jet, given by $g_{\rm m}$ and $k$, the acceleration 
efficiency $g_{\rm acc}$, as well as the spectral index $\beta_L$ of the correlation and the spectral index $a$ of the 
CRs at the sources. 
It turns out, that the bulk of FR-II RGs cannot provide enough HECR power to explain the observed HECR flux, if $Q_{\rm 
cr}<10^{46}\,\text{erg s}^{-1}$ at $L_{151}=10^{26.5}\,\text{W}\,\text{Hz}^{-1}\,\text{sr}^{-1}$ as suggested by the 
most recent correlation models. Here, even a vanishing lepton energy budget $k\ll 1$ and a hard initial CR spectrum 
$a\leq 1$ are not sufficient. In contrast, there is a large variety of different parameter setups that enable a 
significant HECR contribution by FR-I RGs. It is shown for a maximal CR load of the jet, i.e. $g_{\rm m}\sim 
4/7$ and $k=0$, which acceleration efficiency is required dependent on $\beta_L$ and $a$. 

Finally, two proof of principle scenarios are introduced that enable an explanation of the hardening part of the CR 
flux at $10^{18.7}\,\text{eV}\lesssim E \lesssim 10^{19.5}\,\text{eV}$: 
\begin{enumerate}
 \item[(I)] A dominant contribution by FR-I RGs, in the case of a 
low CR load, but a high acceleration efficiency $g_{\rm acc}\gtrsim 0.8$ of these sources. 
However, also a large correlation index $\beta_L\gtrsim 0.9$ is needed, that is disfavored by theoretical 
expectations of the FR-I lobe dynamics \cite{doi:10.1093_mnras_stv2712}. 
 \item[(II)] A dominant contribution by FR-II RGs, in the case of a 
significantly higher CR power of these sources with a vanishing lepton fraction. But such an energetically 
dominant CR population is disfavored by some models \cite{j_1365-2966_2004_08118_x, Croston_2005, Belsole_2007}, that 
suggest $k\gtrsim 1$ in the lobes of FR-II RGs. Nevertheless, such a scenario exhibits some strong implication with 
respect to the whole HECR data: 
Supposing that FR-I RGs provide a rather heavy CR contribution with respect to the FR-II class and an individual, 
close-by FR-I source like Centaurus A provides the observed CRs at energies $\gtrsim 30\,\text{EeV}$ as shown by E+18,  
even the observed spectral behavior of the chemical composition, as well as the arrival directions are likely 
explainable. Further, the additional contribution by individual sources can significantly lower the 
necessary HECR power of FR-II RGs. 
\end{enumerate}
However, the northern hemisphere, as covered by the TA experiment, still misses a luminous, 
close-by FR source that provides the observed CRs above the GZK cut-off energy. 
Hence, further investigations are needed to give a final answer on the contribution of FR RGs to the HECR data. 
\acknowledgments
This work most notably benefits from the development of CRPropa3 and useful discussions with J\"{o}rg Rachen. 
Some of the results in this paper have been derived using the software packages Numpy \cite{vanDerWalt2011}, Pandas 
\cite{mckinney-proc-scipy-2010}, Matplotlib \cite{Hunter:2007}.
\appendix
\section{Details on the inverted simulation setup}
\label{invertedDetails}
In the inverted simulation setup a CR candidate that passes the observer surface needs to stay within the simulation in 
order to enable the observation of candidates with a deflection angle $\theta_i\geq 90\degree$. Here $\theta_i$ denotes 
the angle between the normalized arrival direction $\vec a$ and the normalized source direction $\vec s$. Thus, even 
candidates from the far side with respect to the source in a regular simulation setup can be detected. To account for 
the decreasing detection probability in the case of $\theta_i\rightarrow 90\degree$, the number of detected candidates 
has to be corrected by the factor $|\cos\theta_i|^{-1}$. 

So, even the proper arrival direction can be determined by $\vec{a}_{\rm p} = \mathbf{\Omega}_{\rm rot}\, \vec{a}$ 
using the rotation matrix
$$
\mathbf{\Omega}_{\rm rot}=\begin{pmatrix}
                           \cos\omega+\rho_x^2(1-\cos\omega)  &  \rho_x\rho_y(1-\cos\omega)-\rho_z\sin\omega  &  
\rho_x\rho_z(1-\cos\omega)+\rho_y\sin\omega \\
\rho_x\rho_y(1-\cos\omega)+\rho_z\sin\omega  &  \cos\omega+\rho_y^2(1-\cos\omega)  &  
\rho_y\rho_z(1-\cos\omega)-\rho_x\sin\omega \\
\rho_x\rho_z(1-\cos\omega)-\rho_y\sin\omega  &  \rho_y\rho_z(1-\cos\omega)+\rho_x\sin\omega  &  
\cos\omega+\rho_z^2(1-\cos\omega)
                          \end{pmatrix}\,.
$$
Here, the rotation axis 
$$\vec \rho = \frac{\vec s \times \vec{s}_{\rm p}}{|\vec s \times \vec{s}_{\rm p}|}$$
with the normalized, proper source direction $\vec{s}_{\rm p}$, and the rotation angle 
$\omega=\arccos\left(\vec{s}\cdot 
\vec{s}_{\rm p}\right)$ need to be determined at first. 

\begin{figure}[tbh]
  \centering
    \includegraphics[width=0.49\textwidth]{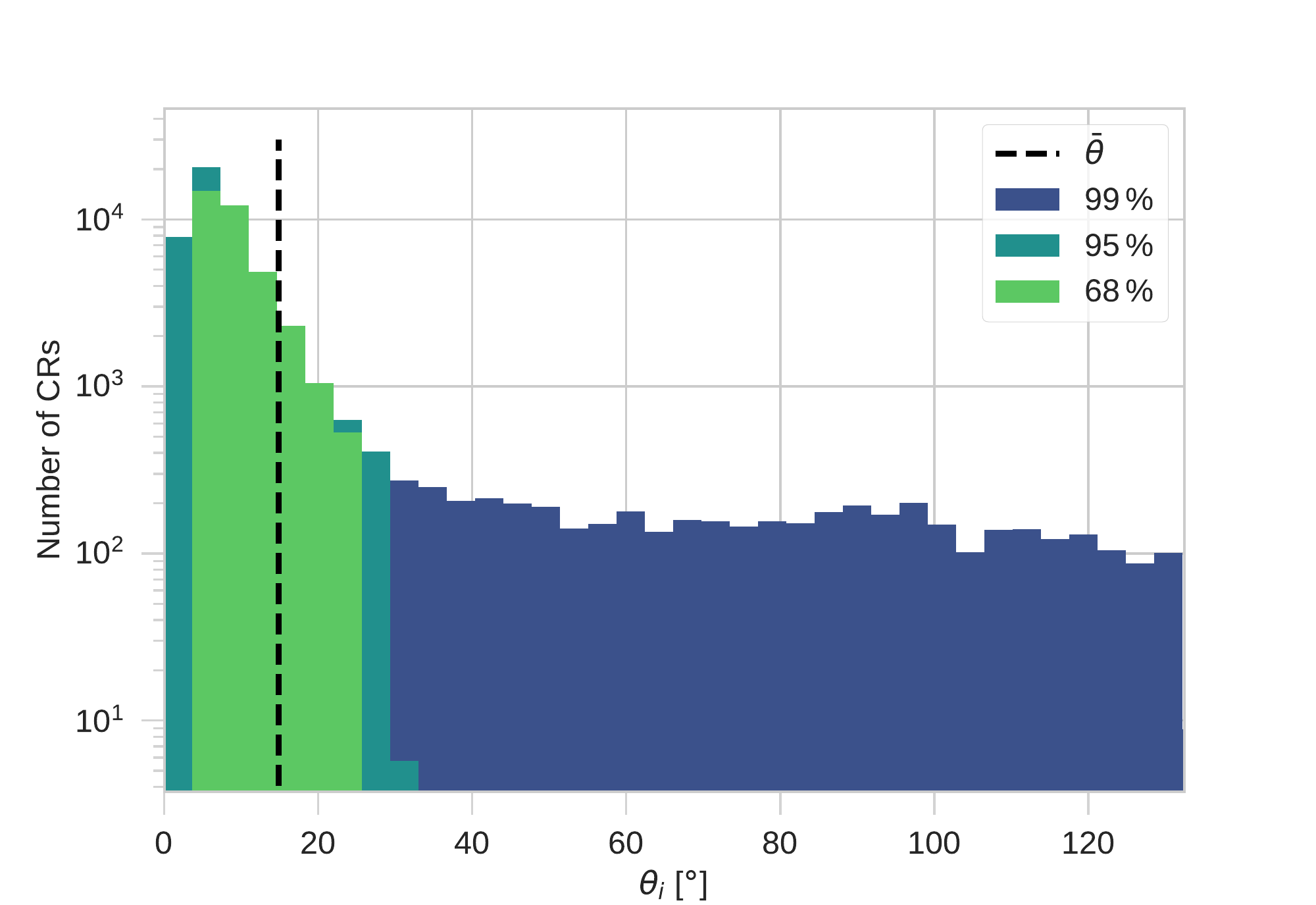}
    \includegraphics[width=0.49\textwidth]{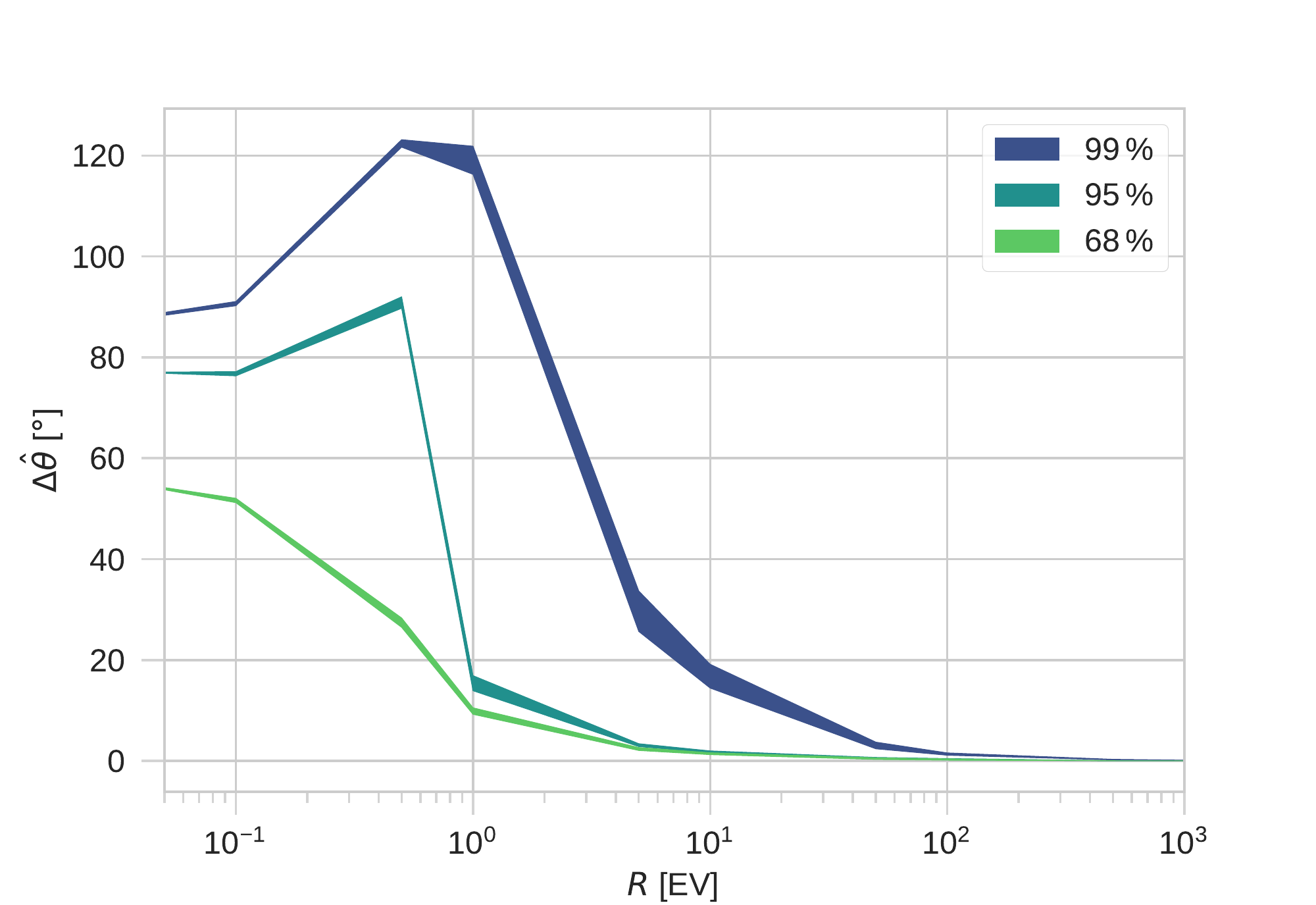}
\caption{Deflections in the D+05 EGMF for a source at a distance of $250\,\text{Mpc}$. \\ \emph{Left:} Different 
percentages of the distribution of $\theta_i$ that are the closest to $\bar{\theta}$ (dashed line) for CRs with a 
rigidity $R=1\,\text{EV}$. \emph{Right:} Maximal absolute deflection error of a given 
percentage of the individual CR candidates. The bands refer to the scattering that results from the effect of 30 
arbitrary source positions.}
\label{invSetup}
\end{figure}

The error of the inverted setup is on the one hand side exposed by the error bands in the Figures \ref{rmsDefl_dolag} 
and \ref{rmsDefl_hack}, which show the standard deviation based on the chosen spatial position of the source. But on the 
other hand, also a given spatial setting causes an uncertainty based on the spread of the resulting distribution of 
deflections, as the inverted simulation setup necessarily provides the sum of all possible deflections dependent on the 
given distance to the source. 
Thus, the absolute error of a chosen setting is given by $\Delta\theta = \sum_i \|\bar{\theta} - \theta_i\|$, where 
$\bar{\theta}$ denotes the mean deflection. Using the extended EGMF structure of D+05 for a source at a distance of 
$250\,\text{Mpc}$, the distribution of $\theta_i$ with respect to $\bar{\theta}$ is analyzed as shown in Fig.\ 
~\ref{invSetup}. Thus, the maximal deflection error $\Delta\hat{\theta}$ for a given percentage of CRs dependent on its 
rigidity is provided as shown in the right Fig.\ ~\ref{invSetup}. Here, the narrow bands indicate, that the chosen 
spatial setting hardly change the resulting maximal deflection error. A small percentage of candidates yields 
$\Delta\hat{\theta}$ of more than $90\degree$ at a few hundreds of PV, that converges towards $90\degree$ with 
decreasing rigidity due to the increase of $\bar{\theta}$. But even at these rigidities, the majority of CRs still 
deviates from $\bar{\theta}$ by less than $\sim 50\degree$. At $1\,\text{EV}$ only about 5\% of the CR candidates 
provide a deflection error of more than about $16\degree$, that continuously decreases to about 5$\degree$ at 
$8\,\text{EV}$. 
Only about 1\% of the sky provides significant deflections errors at the order of several tens of degree at these 
rigidities, which is in good agreement with the extrapolation results by D+05. Thus, a significant over- or 
underestimate of the mean deflections of the HECRs with respect to the proper spatial position of Cygnus A in a regular 
simulation setup is not to be expected.
\bibliographystyle{JHEP} 
\addcontentsline{toc}{section}{Bibliography}
\bibliography{references}

\end{document}